\documentclass[letterpaper,twocolumn,10pt]{article}
\usepackage{usenix2019_v3} %
\usepackage{adjustbox}
\usepackage{amssymb}
\usepackage{amsmath}
\usepackage{multicol}
\usepackage{supertabular}
\usepackage{tikz,graphicx,balance,multirow,xspace,subcaption,longtable}
\usepackage{makecell,amsmath,cleveref,rotating}
\usepackage[T1]{fontenc}
\usepackage{enumitem}
\usepackage[draft,inline,nomargin,index]{fixme}
\usepackage{booktabs}
\usepackage{lastpage}
\usepackage{titling}
\usepackage{xcolor,colortbl}

\usepackage{enumitem}
\setitemize{noitemsep,topsep=0pt,parsep=0pt,partopsep=0pt}

\usepackage[font={small}, textfont=md, skip=.2pt]{caption} %
\newcommand{\tabularscale}{0.80}

\PassOptionsToPackage{hyphens}{url}
\usepackage{hyperref}
\newcolumntype{P}[1]{>{\arraybackslash}p{#1}}
\newcolumntype{X}[1]{>{\centering\arraybackslash}p{#1}}

\expandafter\def\expandafter\UrlBreaks\expandafter{\UrlBreaks%
  \do\a\do\b\do\c\do\d\do\e\do\f\do\g\do\h\do\i\do\j%
  \do\k\do\l\do\m\do\n\do\o\do\p\do\q\do\r\do\s\do\t%
  \do\u\do\v\do\w\do\x\do\y\do\z\do\A\do\B\do\C\do\D%
  \do\E\do\F\do\G\do\H\do\I\do\J\do\K\do\L\do\M\do\N%
  \do\O\do\P\do\Q\do\R\do\S\do\T\do\U\do\V\do\W\do\X%
  \do\Y\do\Z}

\crefformat{section}{\S#2#1#3}
\crefformat{subsection}{\S#2#1#3}
\crefformat{subsubsection}{\S#2#1#3}

\fxsetup{theme=colorsig,mode=multiuser,inlineface=\itshape,envface=\itshape}
\FXRegisterAuthor{ht}{aht}{Hammas}
\FXRegisterAuthor{rn}{arn}{Rishab}
\FXRegisterAuthor{rs}{ars}{Rachee}
\FXRegisterAuthor{pp}{app}{Paul}

\hypersetup{%
  pdftitle={},
  pdfauthor={},
  pdfkeywords={},
  bookmarksnumbered,
  pdfstartview={FitH},
  colorlinks,
  citecolor=black,
  filecolor=black,
  linkcolor=black,
  urlcolor=blue,
  breaklinks=true,
  hypertexnames=false
}

\newcommand\clearrow{\global\let\rowmac\relax}

\newcommand{\para}[1]{{\vspace{.05in} \bf \noindent #1 }}
\newcommand{\parait}[1]{{\vspace{.05in} \em \noindent #1 }}

\renewcommand{\footnotesize}{\scriptsize}
\newcommand{\Sidak}{\v{S}id\'{a}k\ }

\newcommand{\eg}{e.g.,\ }
\newcommand{\etal}{et al.\xspace}
\newcommand{\ie}{i.e.,\ }

\newcommand{\cf}{\emph{cf.,\ }}

\newcommand{\IranRecursionTable}{
\begin{table}[h!]
    \begin{center}
        \begin{tabular}{l|c|c|c} %
                  & \textbf{V4} & \textbf{V6} & \textbf{6to4}         \\
            \hline
            RD    & 10.10.34.35 & d0::11      & 10.10.34.35           \\
            No RD & 10.10.34.35 & d0::11      & \textbf{no injection} \\
        \end{tabular}
    \end{center}
    \caption{The Iranian censor provides distinctive injections to queries sent
        over both v4 and v6, but fails to inject responses for queries to
        resolvers using 6to4 addresses when the query excludes the recursion
        desired flag.}
    \label{tab:iran-recursion-table}
\end{table}
}

\definecolor{red00}{rgb}{1,0.9,0.9}
\definecolor{red0}{rgb}{1,0.8,0.8}
\definecolor{red1}{rgb}{1,0.7,0.7}
\definecolor{red2}{rgb}{1,0.6,0.6}
\definecolor{red3}{rgb}{1,0.5,0.5}
\definecolor{red4}{rgb}{1,0.4,0.4}
\definecolor{red5}{rgb}{1,0.3,0.3}
\definecolor{green0}{rgb}{0.8,1,0.8}
\definecolor{green1}{rgb}{0.7,1,0.7}
\definecolor{green2}{rgb}{0.6,1,0.6}
\definecolor{green3}{rgb}{0.5,1,0.5}
\definecolor{green4}{rgb}{0.4,1,0.4}
\definecolor{green5}{rgb}{0.3,1,0.3}

\newcommand{\FigIndiaCluster}{
    \begin{figure}[t]
     \centering
     \includegraphics[width=0.9\linewidth]{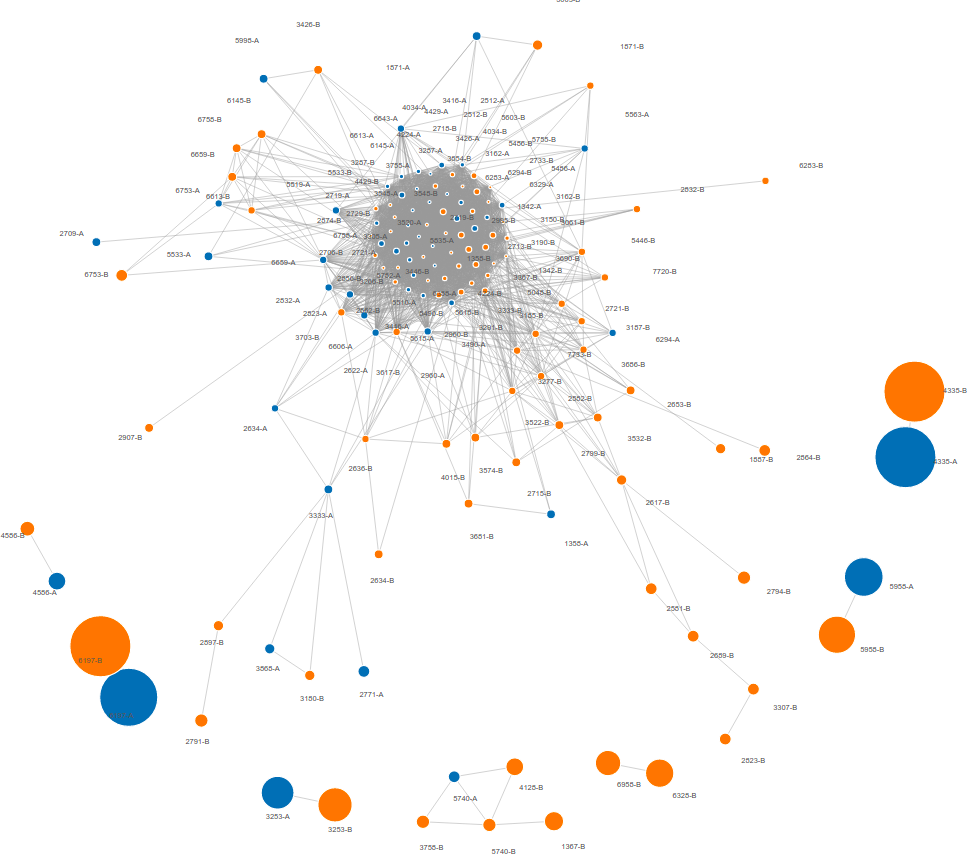}
     \caption{\textbf{India DNS censorship}\,---\,%
            We formed clusters of DNS resolvers using the levenshtein distance between
            the lists resolvers censor. Resolvers that block similar lists are connected,
            with node sizes proportional to the number of distinct domains blocked. We color
            IPv4 resolver interfaces blue, and IPv6 interfaces orange. This example illustrates
            the non-uniformity of India, with some ISPs censoring a large number of domains, and
            many others with small or non-existent block lists.}
    \label{fig:india}
    \end{figure}
}

\begin{document}

\pagestyle{plain} %

\renewcommand{\sectionautorefname}{\S}
\renewcommand{\subsectionautorefname}{\S}
\renewcommand{\subsubsectionautorefname}{\S}
\date{}

\setlength{\droptitle}{-5em}   %
\posttitle{\par\end{center}}   %
\title{{\bf Mind the IP Gap}\\\Large{Measuring the impact of IPv6 on DNS censorship}}
\author{
{\rm Ian Martiny}\\
University of Colorado Boulder
\and
{\rm Hammas Bin Tanveer}\\
University of Iowa
\and 
{\rm Jack Wampler}\\
University of Colorado Boulder
\and
{\rm Rishab Nithyanand}\\
University of Iowa
\and
{\rm Eric Wustrow}\\
University of Colorado Boulder
}

\maketitle

\begin{abstract}
Internet censorship impacts large segments of the Internet, but so far, prior
work has focused almost exclusively on performing measurements using IPv4. As
the Internet grows, and more users connect, IPv6 is increasingly supported and
available to users and servers alike. But despite this steady growth, it remains
unclear if the information control systems that implement censorship (firewalls,
deep packet inspection, DNS injection, etc) are as effective with IPv6 traffic
as they are with IPv4.

In this paper, we perform the first global measurement of DNS censorship on the
IPv6 Internet. Leveraging a recent technique that allows us to discover
IPv6-capable open resolvers (along with their corresponding IPv4 address), we
send over 20~million {\tt A} and {\tt AAAA} DNS requests to DNS resolvers
worldwide, and measure the rate at which they block, at the resolver, network,
and country level as well examine the characteristics of blocked domains.
We observe that while nearly all censors support blocking IPv6, their policies
are inconsistent with and frequently less effective than their IPv4 censorship
infrastructure. Our results suggest that supporting IPv6 censorship is not
all-or-nothing: many censors support it, but poorly.  As a result, these censors
may have to expend additional resources to bring IPv6 censorship up to parity
with IPv4. In the meantime, this affords censorship circumvention researchers a
new opportunity to exploit these differences to evade detection and blocking.
 \end{abstract}

\section{Introduction}\label{sec:intro}

Internet censorship is a global problem that affects over half the world's
population. Censors rely on sophisticated network middleboxes to inspect and
block traffic. A core component of Internet censorship is DNS blocking, and
prior work has extensively studied how DNS censorship occurs, both for specific
countries~\cite{Anonymous2020:TripletCensors,USESEC21:GFWatch} and
globally~\cite{kuhrer2015going,dagon2008corrupted,pearce2017global,scott2016satellite}.
These studies generally perform active measurements of DNS resolvers for large
sets of domains, and identify forged censorship responses from legitimate ones.
Unfortunately, this prior work has focused exclusively on the IPv4 Internet, in
part because scanning the IPv6 Internet for open resolvers is
difficult~\cite{murdock2017target}, owing to its impossible-to-enumerate 128-bit
address space.
In this paper, we perform the first comprehensive global measurement of DNS
censorship on the IPv6 Internet. We leverage a recent network measurement
technique that can discover dual-stack IPv6 open resolvers from their IPv4
counterpart~\cite{hendriks2017potential}, and use these IPv4-IPv6 resolver pairs
to study DNS censorship globally. %
We then use this data to measure the difference in censorship on IPv4 and IPv6.

IPv6 is becoming more widely deployed, with nearly 35\% of current Internet
traffic being served over native IPv6 connections~\cite{Google-IPv6}. However,
IPv6 has fundamentally different performance
characteristics~\cite{Dhamdhere-IMC2012}, network security
policies~\cite{Czyz-NDSS2016}, and network topologies~\cite{Czyz-SIGCOMM2014}
compared to the traditional IPv4 Internet.
While it may seem that censors either do or don't support detecting and
censoring IPv6 DNS in an all-or-nothing fashion, we find that there is a
tremendous range of how well a censor blocks in IPv6 compared to IPv4.
In particular, although nearly all of the countries we study have some support
for IPv6 censorship, we find that most block less effectively in IPv6 compared
to IPv4. For instance, we observe Thailand censors on average 80\% fewer IPv6
DNS resources compared to IPv4 ones, despite a robust nation-wide censorship
system~\cite{gebhart2017internet}.

Studying censorship in IPv6 can provide opportunities for circumvention tools.
By identifying ways that censors miss or incorrectly implement blocking, we can
offer these as techniques that tools can exploit. Moreover, because of the
complex and heterogeneous censorship systems censors operate, many of these
techniques would be costly for censors to prevent, requiring investing
significant resources to close the IPv4/IPv6 gap in their networks. For this
reason, we believe IPv6 can provide unique techniques for circumvention
researchers and tool developers alike, that will be beneficial in the short term
and potentially robust in the longer term.

\medskip
We find a significant global presence of IPv6 DNS censorship --- comparable, but
not identical to well documented IPv4 censorship efforts. Censors demonstrate a
clear bias towards IPv4, censoring \texttt{A} queries in IPv4 at the highest
rates, and a propensity for censoring native record types (\texttt{A} in IPv4,
\texttt{AAAA} in IPv6). At the country level we break down differences by
resolver and domain across resource record and interface type. We find that
multiple countries --- Thailand, Myanmar, Bangladesh, Pakistan, and Iran ---
present consistent discrepancies across all resolvers or domains indicating
centrally coordinated censorship, where the policies that govern IPv4 and IPv6
censorship are managed centrally. Other countries show more varied discrepancies
in the ways that resolvers censor IPv4 and IPv6, due to decentralized models of
censorship, such as that in Russia~\cite{ramesh2020decentralized}, or due to
independent and varied corporate network firewalls.
We also identify behavior indicative of censorship oversight that can be
advantageous to censorship circumvention. For example Brazil and Thailand censor
IPv6 queries that rely on 6to4 bridges at lower rates, presumably due to the
encapsulation of an IPv6 DNS request in an IPv4 packet, instead of appearing as
UDP.

Taken all together, this study provides a first look at IPv6 DNS censorship and
the policy gaps that arise from the IPv6 transition. We provide the following
contributions:

\begin{itemize}
    \item
    We conduct the first large-scale measurement of IPv6 DNS censorship in over
    100~IPv6-connected countries. We find that while most censors support IPv6
    in some capacity, there are significant gaps in how well they censor IPv6.

    \item We provide methodological improvements on measuring DNS censorship
    that avoids relying on cumbersome IP comparisons (that are not
    robust to region-specific DNS nameservers). Our methods are easily
    reproducible, and can be used in future measurement studies.

    \item We characterize the difference in censorship of both network type
    (IPv4 and IPv6), and resource type ({\tt A} and {\tt AAAA} record), and
    identify trends in several countries.

    \item Using our findings, we suggest several new avenues of future
    exploration for censorship circumvention researchers, and censorship
    measurements.

\end{itemize}

The remainder of this paper is organized as follows. \Cref{sec:background} provides
background information in DNS censorship, and the relation of IPv6 to relevant DNS
infrastructure. We outline our compiled methodology and ethical design
considerations in \Cref{sec:methodology} before presenting our findings on the
global prevalence of IPv6 censorship in \Cref{sec:prevalence}. We then dig into
per country analysis based on Resource Record types in \Cref{sec:resources} and
IP protocol version in \Cref{sec:infrastructure}. We select several case studies
to highlight in \Cref{sec:cases} before covering related work in
\Cref{sec:related}. Finally \Cref{sec:discussion} provides discussion and
contextualization of this work before concluding.

\section{Background}\label{sec:background}
Our work is focused on uncovering differences in {\em Internet censorship} that occur
from {\em DNS censorship} mechanisms' failure to effectively adapt to the Internet's
{\em adoption of the IPv6 protocol}. In this section, we provide an overview of
Internet censorship approaches (\Cref{sec:background:censorship}), DNS
censorship mechanisms and infrastructure (\Cref{sec:background:dns}), and the
impact of IPv6 on DNS censorship (\Cref{sec:background:ipv6}).

\subsection{Internet censorship}\label{sec:background:censorship}

\para{What constitutes Internet censorship?}
Internet censorship can be broadly defined as the act of filtering or blocking
access to Internet content \cite{townsend}. Like previous work, we also use
this definition when measuring censorship.

\para{Corporate and national censorship.}
Our working definition of censorship applies to any Internet traffic filtering
or blocking actions, regardless of the entities that perform them or their purpose for doing so. With a few
exceptions, much of prior work has largely focused on measuring censorship with
a presumed attribution of measured events to national content access policies.
However, it is important to note that corporate censorship, where businesses
deploy information controls systems to prevent access to specific types of
content, is also globally prevalent. 
Throughout our work, we take care to identify when it is likely that our
measurements are reporting instances of corporate censorship rather than
national censorship.  This classification is done using Maxmind's Connection
Type database \cite{maxmind-connectiondb} which characterizes IP end-points as
`Corporate', `Cable/DSL' (associated typically with residential networks), or
`Cellular'.
This distinction is important in order to provide context for some of our
results. For example, from our study we find that 0.9\% of our tests in the
United States were censored. Without the additional context that: (1) more that
50\% of all our measurement vantage points in the US were placed in corporate
networks and (2) these end-points were responsible for the identified
censorship events, this finding would be confusing, assumed incorrect, or
incorrectly attributed to a national censorship policy. 

\para{Internet censorship techniques.} 
Blocking access to Internet content can occur in a variety of ways and at
different layers of the networking stack, all of which require the ability to
monitor network traffic passing through the censor's borders. The simplest and
most common approaches are: (1) IP-based blocking in which a censor maintains
blocklists of IP addresses and prevent connections to these IP addresses; (2)
DNS manipulation where censors inject (when the censor is a man-in-the-middle)
or return (when the censor is the resolver) incorrect responses to domains that
are to be censored; and (3) HTTP proxying in which a censor acts as a proxy to
clients within its border with the intention of blocking access to intercepted
`unsuitable' content. 
While these censorship mechanisms are passive and possible to evade, it is
known that countries such as China have deployed comprehensive censorship
infrastructure that is capable of active probing, protocol inspection, and
incorporates multiple approaches for censorship.
In fact, there has been a large body of work to identify censorship techniques
globally
\cite{pearce2017global, niaki2020iclab, scott2016satellite,
sundara2020censored, filasto2012ooni, pearce2017augur, razaghpanah2016exploring}
and specific to individual countries \cite{USESEC21:GFWatch, aryan2013internet,
ramesh2020decentralized, yadav2018light, gebhart2017internet, nabi2013anatomy}.
Unfortunately, the impact of the growing deployment of IPv6 networks has not
been previously studied --- leaving gaps in our knowledge of how censors are
handling the network transition and what opportunities exist for developers of
circumvention tools. Our work fills this gap by studying how the transition to
IPv6 impacts DNS censorship mechanisms.

\subsection{DNS censorship}\label{sec:background:dns}

The Domain Name System (DNS) underpins the global internet by providing
a mapping from human readable hostnames to routable IP addresses making domain
name resolution the first step in almost all connection establishment flows.
However, the widely deployed DNS system is implemented as a plaintext protocol
allowing on-path eavesdroppers to inspect the hostnames as clients attempt to
establish connections and in some cases inject falsified responses to interfere.

Censors have long been known to use DNS injection to block requests, observing
DNS requests and injecting false responses for requests to censored domains.
The Chinese traffic inspection system, called the Great Firewall (GFW), is
documented injecting falsified DNS responses as early as
2002~\cite{global2002great}. This censorship has been shown to be a packet
injection from an on-path adversary monitoring for hostnames in DNS queries
that match regular expressions~\cite{USESEC21:GFWatch}.

\para{Distributed and centralized DNS censorship.}
In a centralized DNS censorship mechanism, a single entity manages the
`blocklists' or filter-rules associated with DNS censorship decisions. This is
the case in countries such as China where traffic inspection devices are housed
at or near border gateways \cite{xu2011internet}. In contrast, countries such
as India and Russia are known to delegate censorship orders to regional
ISPs~\cite{Gosain2017a}
who may choose to implement them either via DNS (typically by reconfiguring
their own resolvers) or other censorship approaches
\cite{ramesh2020decentralized, Yadav2018a, singh2020india}. The latter approach results in
different implementations of DNS censorship and the possibility for
inconsistencies across different ISPs.

\subsection{IPv6} \label{sec:background:ipv6}

\para{IPv6 6to4 bridges.}
As more IPv6-only network connected devices emerge, there is an increasing need
for technologies that facilitate communication betweeb IPv4 and IPv6
end-points. One such technology is 6to4 bridging. The 6to4 protocol was
originally designed as a mechanism to bridge the IPv6 deployment gap as public
6to4 bridges would provide interoperability to legacy systems such that IPv4
services could communicate with IPv6 networks and vice-versa. The protocol
works by assigning a \texttt{/48} subnet with the encoded local address to an
interface then allowing that interface to receive traffic encapsulated in an
IPv4 header at the local address. In this way clients without outgoing
connection to an IPv6 capable network can send IPv6 packets over to an IPv6
capable host who forwards them onward. In our study, we come across many
vantage points which are IPv6 connected via a 6to4 bridge. These are easily
identifiable since their IPv6 addresses are drawn from the \texttt{2002::/16}
address block \cite{RFC3056}.
In the context of DNS, sending a DNS request over IPv6 to
\texttt{2002:0102:0304::} will send an \texttt{[IPv6|UDP|DNS]} packet over IPv6
to the 6to4 gateway at that address which will then encapsulate the packet in
IPv4 to 1.2.3.4 (whose 32-bit address is encoded in the IPv6 address
\texttt{0102:0304}) in order to transit a portion of the network with no
connectivity. For this portion of the journey the packet is
\texttt{[IPv4|IPv6|UDP|DNS]} until it reaches the destination or another
network capable of routing IPv6 packets.

\para{The impact of IPv6 on DNS censorship.}
As with other critical protocols, DNS has also adapted to the IPv6 protocol.
The {\tt AAAA} resource record type was introduced to aid in the resolution of
domains to their IPv6 addresses. Further, the existing DNS protocol is
IP-independent and can therefore be deployed on IPv4 and IPv6 networks.
Therefore, it is now possible and common for IPv4 and IPv6-hosted DNS servers
to receive both {\tt A} and {\tt AAAA} queries. This is in contrast to
a IPv4-dominant Internet where {\tt A} queries to IPv4 resolvers were the norm.
The changes outlined above also influence the mechanics and success of DNS
censorship operations. A theoretically comprehensive DNS censorship strategy
using response injection requires traffic monitoring infrastructure to: (1)
analyze both IPv4 and IPv6 traffic and (2) parse both \texttt{A} and
\texttt{AAAA} queries accounting for hostnames that may not implement resource
records of one type or the other. 
Our work provides a snapshot documentation of contemporary
censorship strategies through the IPv6 transition.

\section{Dataset and Methodology}\label{sec:methodology}

The analysis presented in the remainder of this paper is based on the results of
22.4M DNS {\tt A} and {\tt AAAA} resolution requests for 714 domains sent to
7,843 IPv4- and IPv6-capable resolvers located in 128 countries.
In this section, we explain our process for identifying resolver targets for
our queries (\Cref{sec:methodology:resolvers}), domains that are the subject of
our queries (\Cref{sec:methodology:domains}), and our process for identifying
the occurrence of a censorship event from the results of each query
(\Cref{sec:methodology:censorship}).

\subsection{Selecting resolvers} \label{sec:methodology:resolvers}
Our work is aimed at characterizing the inconsistencies that exist in the
handling of IPv4- and IPv6-related DNS queries --- \ie differences in the
handling of {\tt A} and {\tt AAAA} query types over IPv4 and IPv6
connections. Therefore, we required that each resolver used for our
measurements was IPv4- and IPv6-capable.

\para{Identifying resolvers with IPv4- and IPv6-capabilities.}
Our approach, builds on the work of Hendricks \etal
\cite{hendriks2017potential} who identified IPv6 open resolvers to measure the
potential for IPv6-based DDoS attacks.

\parait{IPv6-only domain.}
We begin by creating a new domain, owned and controlled by us, with the name server
set to an IPv6-only record. In other words, to query this domain (which we call
the IPv6-only NS domain), one must contact the IPv6-only name server.

\parait{Identifying IPv4-capable resolvers.}
We use {\tt zmap} \cite{Durumeric13zmap} to scan the entire IPv4 address
space and issue a DNS {\tt A} query on port 53 for a separate IPv4 control
domain we also control. This yields an initial list of 7.2M IPv4 DNS resolvers.

\parait{Verifying IPv6 capabilities of IPv4-capable resolvers.}
Next, we issue a DNS {\tt A} query for a resolver-specific subdomain of our
IPv6-only NS domain to each of these IPv4 DNS resolvers. The subdomain encodes
the IPv4 address of the resolver being targeted. Therefore, if our domain was
{\tt v6onlyNS.io} and our resolver target was {\tt 1.1.1.1}, our DNS query
requested the {\tt A} record for {\tt 1-1-1-1.v6onlyNS.io}.
Since IPv4-only resolvers will not be able to communicate with our IPv6-only
Name Server, we expect this resolution to fail for IPv4-only resolvers.
On the other hand, resolvers with any form of IPv6 connectivity will be
able to connect to our IPv6-only Name Server.
Thus, by examining the logs of our IPv6-only name server, we are able
to identify the set of resolvers that successfully reached our server and
their corresponding IPv4 addresses. The associated IPv6 address for each
successful query is extracted from packet captures of the IPv6-only name
server giving us 33.7K (IPv4, IPv6) address matchings.

\para{Filtering and geolocating resolvers.}
The approach detailed above yields matchings that suggest the presence of
`infrastructure' resolvers --- \eg multiple IPv4 resolvers have the same IPv6
address associated with them. These are cases where the IPv4 resolver simply
forwards requests to a dedicated multi-machine DNS infrastructure rather than
performing the resolution by itself. Although this does not change the validity
of our results regarding the IPv6-related inconsistencies of resolvers, we
still remove these cases in order to minimize the influence of such
infrastructure resolvers. This yields 14.9K resolver pairs.
Finally, to confirm the correctness of our list of IPv4/IPv6 resolver pairs, we:
(1) use the Maxmind GeoIP dataset \cite{maxmind-connectiondb} to geolocate the
IPv4 and IPv6 addresses of a resolver pair and keep those which belong to the
same region which reduces us to 14.2K resolver pairs, (2) perform a follow up
scan by issuing {\tt A} and {\tt AAAA} requests for both our control domains
using {\tt zdns} \cite{Durumeric13zmap} and filter out those pairs where an
incorrect response was received.
In total, we obtained 7,843 resolver pairs across 106 different countries.
\Cref{tab:appendix:rates} illustrates the geographic distribution of the
resolvers.

\subsection{Selecting target domains}
\label{sec:methodology:domains}
In order to identify inconsistencies in IPv4/IPv6-related DNS censorship we
require that the domains we use are: (1) sensitive and likely to be censored in
a large number of countries and (2) have valid {\tt A} and {\tt AAAA} records
associated with them.

\para{Identifying sensitive domains.} Censored Planet's `Satellite' project,
which performs global longitudinal measurements of DNS censorship,
\cite{sundara2020censored} maintains a list of domains that combines sensitive
domains in each country with popular domains randomly chosen from the Alexa
Top-10K. The sensitive domains in this list were gathered by the Citizen Lab
using regional experts to curate lists for each country
\cite{citizenlab-blocklists}. Given the input from experts and the availability
of comparable validation data from Censored Planet, we utilize this list as the
starting point for our study as well. We start with 2506 sensitive domains.

\para{Identifying usable domains.} Not all the domains on the Satellite list
are usable in our study since they do not have IPv6 connectivity or {\tt AAAA}
records.
We filter out unusable domains by using {\tt zdns} to perform {\tt A} and {\tt
AAAA} resource record requests from Google and Cloudflare's four public DNS
resolvers ({\tt 8.8.8.8}, {\tt 8.8.4.4}, {\tt 1.1.1.1}, and {\tt 1.0.0.1}) and
removing those with invalid {\tt A} or {\tt AAAA} records. A record is invalid
if it does not contain valid resource appropriate IPv4 or IPv6 addresses. This
gives us with 775 sensitive and infrastructure appropriate domains.
Finally, we follow-up by making TLS connections (using {\tt zgrab2}), from our
uncensored vantage point, to each of the IP addresses contained in the DNS
responses. We locally verify the obtained TLS certificates and exclude all
domains whose verification fails. 
This final list contains 714 sensitive domains whose TLS certificates and IPv4
and IPv6 addresses are valid. We use this list for all the measurements
reported in this paper.

\subsection{Identifying DNS censorship events} \label{sec:methodology:censorship}
In general, our goal is to err on the side of caution and avoid false-positives
in our censorship determination. We achieve this by accounting for
unreliability of resolvers and domains in our lists.

\para{Removing unstable resolvers.}
For each of the 7,843 (IPv4, IPv6) resolver pairs and 714 sensitive domains, we
send a DNS {\tt A} and {\tt AAAA} resource record request for a domain to the
IPv4 and IPv6 addresses associated with the resolver. We follow this up with
a {\tt A} and {\tt AAAA} resource record request for a set of 3 control domains
(owned and operated by us).
We discard data from pairs which failed to resolve any one of our control
domains correctly since this is a sign of resolver instability. This left us
with 7,441 stable resolver pairs.

\para{Distinguishing censorship from domain instability.}
For the remaining resolvers, we extract the IPv4 and IPv6 resource records
returned for each domain --- even those that arrive from multiple responses to
a single query (a sign of an on-path censor). Next, we use \texttt{zgrab2} to
establish TLS connections to the IP addresses contained in the resource
records. In each of these connections, we set the TLS SNI to be the domain
whose records were requested. We then locally verify the validity of the
retrieved TLS certificates.
Since the domains themselves might be unreliable, we repeat this verification
procedure three times. Only if this step fails all three times do we conclude
that the IP we obtained from that resolver for that domain was censored. 

\subsection{Ethics}\label{sec:methodology:ethics}
Our experimental design has incorporated ethical considerations into the
decision-making process at multiple stages.
Censorship measurement has inherent risks and trade-offs: better understanding
of censorship can help support and inform users, but specific measurements may
carry risk to participants or network users.
We rely on The Menlo Report~\cite{menlo}, its companion
guide~\cite{menlo-companion}, and the censorship specific ethical measurement
guidelines discussed by Jones \etal \cite{jones2015ethical} to 
carefully weigh these trade-offs in our experimental design.

\para{Consent.}
To align with the guiding principle of \textit{respect for persons} we
structure the data collection to implicate as few individuals as possible.
Specifically we rely on open resolvers which typically have little or no direct
association with individuals in lieu of measurement from client based
software. While we cannot acquire direct or proxy consent from the operators of
the open resolvers we consider the trade-off between the implied consent
standard and the value in the measurements we make. 
We note that the goal of our ethical analysis is not to eliminate risk, but to
minimize it wherever possible. As noted by Jones \etal in some cases acquiring
consent from operators may not only be impossible, but could increase the risk to
operators as it introduces their acknowledgement of, or active participation
in, the measurement at hand~\cite{jones2015ethical}. This analysis aligns with
previous work relying on open resolvers to collect impactful results while
minimizing risk on
individuals~\cite{pearce2017global,scott2016satellite,sundara2020censored}.

\para{Privacy.}
Our study collects no personal data about any end~users or open
resolver operators. The analysis completed herein uses resolver addresses,
public Anonymous System (AS) identifiers, and country codes.  All measurements
are initiated from within the United States.
Beyond this,
measurement domains are not drawn from any human browsing patterns or history
as the suspected censored domains are a subset of the Satellite measurement
results (\cf \Cref{sec:methodology:domains}).

\para{Resource usage.} 
The vantage that was used for data collection is connected to the internet with
a 1 Gbps interface that scanned using the default rates for {\tt zmap} and {\tt
zgrab2} tools (line rate). However, the structure of the scan was established
such that individual resolvers and domains for the DNS probe and TLS
certificate validation respectively would be accessed in round robin order ---
\ie when probing the open resolvers every target would receive a first
request before any target would receive the subsequent request.  Equivalently
for validating TLS certificates, each of the 714 target domains would receive
a first attempted handshake before any target would receive a subsequent
handshake, limiting the bandwidth any one host will receive.
\section{Prevalence of DNS Censorship}
\label{sec:prevalence}
\para{Overview.} In this section, we focus on \emph{providing a high-level
understanding on the prevalence of DNS censorship on IPv4 and IPv6 networks}.
Specifically, we measure the global prevalence of DNS censorship that occur in
the following four cases: 
(1) a DNS {\tt A} query is sent over IPv4, 
(2) a DNS {\tt AAAA} query is sent over IPv4, 
(3) a DNS {\tt A} query is sent over IPv6, and 
(4) a DNS {\tt AAAA} query is sent over IPv6.
While much of prior work has focused on case (1), the increased adoption of
IPv6 necessitates the analysis of cases (2-4) which are provided in our work.
In each case, we use our collected dataset (\cf \Cref{sec:methodology}) to
summarize the base rate of censorship.

\begin{table}[t]
  \centering
  \small
  \scalebox{\tabularscale}{
  \begin{tabular}{lcccccc}
    \toprule
    {\bf Country} &  {\bf Resolver} & {\bf IPv4} & {\bf IPv4} & {\bf IPv6} & {\bf IPv6} & {\bf Avg.} \\
    {}  & {\bf pairs} & {\tt A} & {\tt AAAA}  & {\tt A} & {\tt AAAA} & {}\\
    \midrule
    China (CN) & \color{black} 194 & {\cellcolor[HTML]{E1EDF8}}
    \color[HTML]{000000} \color{black} 29.27 & {\cellcolor[HTML]{6CAED6}}
    \color[HTML]{F1F1F1} \color{black} 32.32 & {\cellcolor[HTML]{F7FBFF}}
    \color[HTML]{000000} \color{black} 28.41 & {\cellcolor[HTML]{77B5D9}}
    \color[HTML]{000000} \color{black} 32.08 & \color{black} 30.52 \\
    Iran (IR) & \color{black} 277 & {\cellcolor[HTML]{6AAED6}}
    \color[HTML]{F1F1F1}
    \color{black} 25.12 & {\cellcolor[HTML]{8DC1DD}} \color[HTML]{000000}
    \color{black} 24.49 & {\cellcolor[HTML]{EAF2FB}} \color[HTML]{000000}
    \color{black} 21.95 & {\cellcolor[HTML]{F7FBFF}} \color[HTML]{000000}
    \color{black} 21.45 & \color{black} 23.25 \\
    Hong Kong (HK) & \color{black} 67 & {\cellcolor[HTML]{6CAED6}}
    \color[HTML]{F1F1F1} \color{black} 7.00 & {\cellcolor[HTML]{D8E7F5}}
    \color[HTML]{000000} \color{black} 5.57 & {\cellcolor[HTML]{F7FBFF}}
    \color[HTML]{000000} \color{black} 4.91 & {\cellcolor[HTML]{E7F1FA}}
    \color[HTML]{000000} \color{black} 5.24 & \color{black} 5.68 \\
    Russia (RU) & \color{black} 312 & {\cellcolor[HTML]{6AAED6}}
    \color[HTML]{F1F1F1} \color{black} 5.51 & {\cellcolor[HTML]{DCE9F6}}
    \color[HTML]{000000} \color{black} 4.78 & {\cellcolor[HTML]{F7FBFF}}
    \color[HTML]{000000} \color{black} 4.49 & {\cellcolor[HTML]{F7FBFF}}
    \color[HTML]{000000} \color{black} 4.50 & \color{black} 4.82 \\
    Ukraine (UA) & \color{black} 35 & {\cellcolor[HTML]{6CAED6}}
    \color[HTML]{F1F1F1} \color{black} 6.49 & {\cellcolor[HTML]{EDF4FC}}
    \color[HTML]{000000} \color{black} 3.05 & {\cellcolor[HTML]{F7FBFF}}
    \color[HTML]{000000} \color{black} 2.64 & {\cellcolor[HTML]{F2F7FD}}
    \color[HTML]{000000} \color{black} 2.87 & \color{black} 3.76 \\
    Indonesia (ID) & \color{black} 56 & {\cellcolor[HTML]{6AAED6}}
    \color[HTML]{F1F1F1} \color{black} 6.46 & {\cellcolor[HTML]{E6F0F9}}
    \color[HTML]{000000} \color{black} 2.99 & {\cellcolor[HTML]{F4F9FE}}
    \color[HTML]{000000} \color{black} 2.41 & {\cellcolor[HTML]{F7FBFF}}
    \color[HTML]{000000} \color{black} 2.26 & \color{black} 3.53 \\
    Argentina (AR) & \color{black} 47 & {\cellcolor[HTML]{6CAED6}}
    \color[HTML]{F1F1F1} \color{black} 5.20 & {\cellcolor[HTML]{BCD7EB}}
    \color[HTML]{000000} \color{black} 3.51 & {\cellcolor[HTML]{DFECF7}}
    \color[HTML]{000000} \color{black} 2.22 & {\cellcolor[HTML]{F7FBFF}}
    \color[HTML]{000000} \color{black} 1.28 & \color{black} 3.05 \\
    Thailand (TH) & \color{black} 186 & {\cellcolor[HTML]{6AAED6}}
    \color[HTML]{F1F1F1} \color{black} 8.25 & {\cellcolor[HTML]{F4F9FE}}
    \color[HTML]{000000} \color{black} 1.18 & {\cellcolor[HTML]{F5F9FE}}
    \color[HTML]{000000} \color{black} 1.13 & {\cellcolor[HTML]{F7FBFF}}
    \color[HTML]{000000} \color{black} 0.93 & \color{black} 2.87 \\
    Malaysia (MY) & \color{black} 50 & {\cellcolor[HTML]{6AAED6}}
    \color[HTML]{F1F1F1} \color{black} 4.92 & {\cellcolor[HTML]{E3EEF8}}
    \color[HTML]{000000} \color{black} 2.03 & {\cellcolor[HTML]{F7FBFF}}
    \color[HTML]{000000} \color{black} 1.27 & {\cellcolor[HTML]{F5FAFE}}
    \color[HTML]{000000} \color{black} 1.35 & \color{black} 2.39 \\
    Mexico (MX) & \color{black} 150 & {\cellcolor[HTML]{6AAED6}}
    \color[HTML]{F1F1F1} \color{black} 3.78 & {\cellcolor[HTML]{E9F2FA}}
    \color[HTML]{000000} \color{black} 2.05 & {\cellcolor[HTML]{F7FBFF}}
    \color[HTML]{000000} \color{black} 1.75 & {\cellcolor[HTML]{EEF5FC}}
    \color[HTML]{000000} \color{black} 1.94 & \color{black} 2.38 \\
    Bangladesh (BD) & \color{black} 29 & {\cellcolor[HTML]{6AAED6}}
    \color[HTML]{F1F1F1} \color{black} 6.42 & {\cellcolor[HTML]{EFF6FC}}
    \color[HTML]{000000} \color{black} 1.28 & {\cellcolor[HTML]{F6FAFF}}
    \color[HTML]{000000} \color{black} 0.89 & {\cellcolor[HTML]{F7FBFF}}
    \color[HTML]{000000} \color{black} 0.81 & \color{black} 2.35 \\
    Colombia (CO) & \color{black} 27 & {\cellcolor[HTML]{6AAED6}}
    \color[HTML]{F1F1F1} \color{black} 3.39 & {\cellcolor[HTML]{75B4D8}}
    \color[HTML]{000000} \color{black} 3.28 & {\cellcolor[HTML]{EAF2FB}}
    \color[HTML]{000000} \color{black} 1.45 & {\cellcolor[HTML]{F7FBFF}}
    \color[HTML]{000000} \color{black} 1.14 & \color{black} 2.31 \\
    Italy (IT) & \color{black} 38 & {\cellcolor[HTML]{B7D4EA}}
    \color[HTML]{000000}
    \color{black} 2.36 & {\cellcolor[HTML]{F7FBFF}} \color[HTML]{000000}
    \color{black} 2.10 & {\cellcolor[HTML]{D6E5F4}} \color[HTML]{000000}
    \color{black} 2.25 & {\cellcolor[HTML]{6AAED6}} \color[HTML]{F1F1F1}
    \color{black} 2.52 & \color{black} 2.31 \\
    Brazil (BR) & \color{black} 160 & {\cellcolor[HTML]{6AAED6}}
    \color[HTML]{F1F1F1} \color{black} 2.67 & {\cellcolor[HTML]{F7FBFF}}
    \color[HTML]{000000} \color{black} 1.67 & {\cellcolor[HTML]{D8E7F5}}
    \color[HTML]{000000} \color{black} 1.99 & {\cellcolor[HTML]{CFE1F2}}
    \color[HTML]{000000} \color{black} 2.08 & \color{black} 2.10 \\
    Bulgaria (BG) & \color{black} 30 & {\cellcolor[HTML]{6AAED6}}
    \color[HTML]{F1F1F1} \color{black} 3.24 & {\cellcolor[HTML]{89BEDC}}
    \color[HTML]{000000} \color{black} 2.91 & {\cellcolor[HTML]{F4F9FE}}
    \color[HTML]{000000} \color{black} 1.15 & {\cellcolor[HTML]{F7FBFF}}
    \color[HTML]{000000} \color{black} 1.06 & \color{black} 2.09 \\
    Poland (PL) & \color{black} 48 & {\cellcolor[HTML]{C1D9ED}}
    \color[HTML]{000000} \color{black} 2.00 & {\cellcolor[HTML]{EEF5FC}}
    \color[HTML]{000000} \color{black} 1.13 & {\cellcolor[HTML]{6AAED6}}
    \color[HTML]{F1F1F1} \color{black} 2.90 & {\cellcolor[HTML]{F7FBFF}}
    \color[HTML]{000000} \color{black} 0.96 & \color{black} 1.75 \\
    South Africa (ZA) & \color{black} 93 & {\cellcolor[HTML]{6AAED6}}
    \color[HTML]{F1F1F1} \color{black} 2.41 & {\cellcolor[HTML]{D5E5F4}}
    \color[HTML]{000000} \color{black} 1.60 & {\cellcolor[HTML]{F7FBFF}}
    \color[HTML]{000000} \color{black} 1.18 & {\cellcolor[HTML]{EBF3FB}}
    \color[HTML]{000000} \color{black} 1.33 & \color{black} 1.63 \\
    Korea (KR) & \color{black} 632 & {\cellcolor[HTML]{6AAED6}}
    \color[HTML]{F1F1F1} \color{black} 2.33 & {\cellcolor[HTML]{F7FBFF}}
    \color[HTML]{000000} \color{black} 1.06 & {\cellcolor[HTML]{E8F1FA}}
    \color[HTML]{000000} \color{black} 1.24 & {\cellcolor[HTML]{EAF3FB}}
    \color[HTML]{000000} \color{black} 1.22 & \color{black} 1.46 \\
    Chile (CL) & \color{black} 65 & {\cellcolor[HTML]{6CAED6}}
    \color[HTML]{F1F1F1}
    \color{black} 2.67 & {\cellcolor[HTML]{F7FBFF}} \color[HTML]{000000}
    \color{black} 0.70 & {\cellcolor[HTML]{E3EEF9}} \color[HTML]{000000}
    \color{black} 1.10 & {\cellcolor[HTML]{F3F8FE}} \color[HTML]{000000}
    \color{black} 0.79 & \color{black} 1.32 \\
    Romania (RO) & \color{black} 44 & {\cellcolor[HTML]{6AAED6}}
    \color[HTML]{F1F1F1} \color{black} 2.30 & {\cellcolor[HTML]{EAF2FB}}
    \color[HTML]{000000} \color{black} 1.03 & {\cellcolor[HTML]{F7FBFF}}
    \color[HTML]{000000} \color{black} 0.83 & {\cellcolor[HTML]{F1F7FD}}
    \color[HTML]{000000} \color{black} 0.93 & \color{black} 1.27 \\
    Spain (ES) & \color{black} 49 & {\cellcolor[HTML]{EAF2FB}} \color[HTML]{000000}
    \color{black} 0.91 & {\cellcolor[HTML]{F7FBFF}} \color[HTML]{000000}
    \color{black} 0.75 & {\cellcolor[HTML]{BFD8ED}} \color[HTML]{000000}
    \color{black} 1.40 & {\cellcolor[HTML]{6AAED6}} \color[HTML]{F1F1F1}
    \color{black} 1.94 & \color{black} 1.25 \\
    India (IN) & \color{black} 226 & {\cellcolor[HTML]{C8DCF0}}
    \color[HTML]{000000} \color{black} 1.27 & {\cellcolor[HTML]{6AAED6}}
    \color[HTML]{F1F1F1} \color{black} 1.52 & {\cellcolor[HTML]{F7FBFF}}
    \color[HTML]{000000} \color{black} 1.04 & {\cellcolor[HTML]{DFEBF7}}
    \color[HTML]{000000} \color{black} 1.16 & \color{black} 1.25 \\
    Belgium (BE) & \color{black} 31 & {\cellcolor[HTML]{6CAED6}}
    \color[HTML]{F1F1F1} \color{black} 1.56 & {\cellcolor[HTML]{C6DBEF}}
    \color[HTML]{000000} \color{black} 1.26 & {\cellcolor[HTML]{F7FBFF}}
    \color[HTML]{000000} \color{black} 0.95 & {\cellcolor[HTML]{F7FBFF}}
    \color[HTML]{000000} \color{black} 0.95 & \color{black} 1.18 \\
    Turkey (TR) & \color{black} 114 & {\cellcolor[HTML]{6CAED6}}
    \color[HTML]{F1F1F1} \color{black} 1.29 & {\cellcolor[HTML]{F7FBFF}}
    \color[HTML]{000000} \color{black} 0.96 & {\cellcolor[HTML]{7AB6D9}}
    \color[HTML]{000000} \color{black} 1.27 & {\cellcolor[HTML]{E3EEF9}}
    \color[HTML]{000000} \color{black} 1.02 & \color{black} 1.14 \\
    Viet Nam (VN) & \color{black} 252 & {\cellcolor[HTML]{6AAED6}}
    \color[HTML]{F1F1F1} \color{black} 1.53 & {\cellcolor[HTML]{DFEBF7}}
    \color[HTML]{000000} \color{black} 0.89 & {\cellcolor[HTML]{87BDDC}}
    \color[HTML]{000000} \color{black} 1.42 & {\cellcolor[HTML]{F7FBFF}}
    \color[HTML]{000000} \color{black} 0.67 & \color{black} 1.13 \\

    \midrule
    {\bf Global} & \color{black} 7,428 & {\cellcolor[HTML]{6AAED6}}
    \color[HTML]{F1F1F1} \color{black} 3.10 & {\cellcolor[HTML]{DCE9F6}}
    \color[HTML]{000000} \color{black} 1.83 & {\cellcolor[HTML]{F7FBFF}}
    \color[HTML]{000000} \color{black} 1.77 & {\cellcolor[HTML]{F7FBFF}}
    \color[HTML]{000000} \color{black} 1.60 & \color{black} 2.07 \\
    
  \bottomrule
  \end{tabular}
  }
  \caption{
  Top-25 countries with over 25 resolver pairs and the highest average
  rates of DNS censorship across both query types and network interfaces. Rates
  are expressed as the percentage of censored DNS queries.
  Darker shaded cells indicate a higher rate of DNS censorship (compared to the
  country's average).  %
  The average rate of censorship for a country is computed across all four
  IP/query combinations.
  The global row contains the mean of each column and includes data from the
  countries with less than 25 resolvers. These means weigh the contribution of
  each country equally, rather than weighted by the number of resolvers used in
  tests.   }
  \label{tab:prevalence:rates}
\end{table}

\para{How common is DNS censorship?} 
Our data shows the mean base rate of DNS censorship among the 106 countries
included in our study, across all query and network types, for our list of
domains is 2.1\%. 
A snippet of the base rates observed in each of our four {\tt A/AAAA}-IPv4/IPv6
combinations for the 25 countries which have at least 25 pairs of resolvers that
were tested and perform the most censorship is illustrated in
\Cref{tab:prevalence:rates}. A full description along with breakdowns by
network connection types is provided in the Appendix (\Cref{tab:appendix:rates}
and \Cref{tab:prevalence:rates:broken}).
In general, our results concur with prior work which has also found high levels
of DNS censorship in China, Iran, Russia, and Hong Kong. As one might expect,
when we break down our results by censorship in corporate and non-corporate
networks (\Cref{tab:prevalence:rates:broken}), we see that the presence of
countries including Spain, Australia, Korea, and Japan are primarily due to
corporate censorship. 

\para{Trends in censorship of IPv6-related queries in heavily censoring
countries.}
By measuring IPv6-related behaviors of censorship mechanisms, we uncover
a large number of DNS censorship inconsistencies in heavily censoring
countries.
Because our dataset is balanced (\ie all the domains have both {\tt A} and
{\tt AAAA} records and all the tested resolvers have an IPv4 and IPv6
interface), if censorship is independent of the query type and interface, we
expect to see a uniform rate of censorship across all query types and interface
combinations in \Cref{tab:prevalence:rates}. 
However, we find that this is not true. We observe two trends common to many of
the countries performing the most censorship. 
First, we see that, in comparison to any other query-interface combination,
{\em {\tt A} queries sent over IPv4 are the most heavily censored}. Second, we
see that {\em{\tt AAAA} queries are censored less than {\tt A} queries,
regardless of whether they are sent over IPv4 and IPv6 networks}.
This immediately suggests that censorship apparatus in a large number of
censoring regions are not fully IPv6-capable. Besides the possibility of
misconfiguration of the DNS censorship mechanisms, this may also be because
censors in these regions are not yet widely deployed on IPv6 networks in their
country.
The only exception to both these generalizations is China --- the most
censoring country in our data. China shows an unusual {\em preference towards
blocking {\tt AAAA} records regardless of whether they are sent over IPv4 or
IPv6}. We explore China and several other countries with
interesting patterns as specific case studies in \Cref{sec:cases}.

\section{Censorship of Resource Records} \label{sec:resources}

\para{Overview.}
In this section, we focus on {\it identifying and characterizing differences in
the handling of IPv4 and IPv6 resource records} in DNS censorship deployments.
Specifically, we seek to answer the following questions: 
(\Cref{sec:resources:country}) In which countries is the censorship of IPv4
\textbf{resource records} (DNS {\tt A} queries) significantly different than the
censorship of IPv6 resource records (DNS {\tt AAAA} queries)?,
(\Cref{sec:resources:resolvers}) what are the characteristics of the
\textbf{resolvers} which exhibit differences in the handling of {\tt A} and {\tt
AAAA} queries?, and 
(\Cref{sec:resources:domains}) what are the characteristics of \textbf{domains}
in which these differences are frequently observed?
\subsection{{\tt A} vs. {\tt AAAA} resource censorship} \label{sec:resources:country}
We use the responses received from our {\tt A} and {\tt AAAA} queries sent to
the same set of resolvers and for the same set of domains (\cf
\Cref{sec:methodology} for data collection methodology). We then apply the
censorship determination methods described in \Cref{sec:methodology:censorship}
to measure the prevalence of censorship on our {\tt A} and {\tt AAAA} DNS
queries. Finally, we perform statistical tests to identify significant
differences in the prevalence of censorship of {\tt A} and {\tt AAAA} queries
within each country.

\para{Identifying differences within a country.}
To measure differences in DNS query handling within a specific country, we
compare the prevalence of censorship on {\tt A} and {\tt AAAA} queries by
aggregating responses across {each resolver within the country}. This presents
us with two distributions (one each for the group of {\tt A} and {\tt AAAA}
queries) of the fraction of censored domains observed at each resolver in the
country.
We use a two-sample $t$-test to verify statistical significance of any observed
differences between the two groups for each country. In our statistical
analysis, we aim to achieve a significance level of 5\% ($p \leq$  .05)
\emph{over all our findings}. Therefore, we apply a \Sidak correction
\cite{abdi2007bonferroni} to control for Type I (false-positive) errors from
multiple hypothesis testing. 
This requires $p \leq 1-{.05}^{1/n_{c}}$ for classifying a difference as
significant, where $n_c$ is the total number of countries in our dataset
(106). This approach reduces the likelihood of false-positive reports
of within-country differences.
The presence of a statistically significant difference for a specific country
would imply that \texttt{A} and \texttt{AAAA} resource types appear to undergo
different censorship mechanisms within that country (if a centralized mechanism
for censorship exists) or that a significant number of resolvers within that
country have inconsistencies in their censoring of each query type.
A summary of our results are presented in \Cref{tab:resources:countries}. 

\para{How many countries demonstrate large-scale inconsistencies in their
handling of {\tt A} and {\tt AAAA} queries?} 
In total, only seven countries showed a statistically significant difference
in the rate at which {\tt A} and {\tt AAAA} DNS requests were blocked. We note
that this is a conservative lower-bound due to the statistical test used, which
minimizes false positive errors at the expense of false negatives.
This finding suggests the presence of independent censorship mechanisms for
handling each query type in the seven countries (Thailand, Bangladesh,
Pakistan, Chile, Vietnam, Korea, and China).
Of these, six (China being the only exception) were found to have lower
blocking rates for {\tt AAAA} queries than {\tt A} queries. In fact, the {\tt
AAAA} censorship rates were between 36-78\% lower than the {\tt A} censorship
rate suggesting that their censorship mechanisms for {\tt AAAA} queries that
are associated with IPv6 connectivity are still lagging. 
Further analysis shows that the differences are mostly found on the IPv4
interfaces of our resolvers (\cf {\it IPv4 resolvers} column in
\Cref{tab:resources:countries}) where the {\tt AAAA} censorship rates were up to
86\% lower than the {\tt A} censorship rates. 
This finding is indicative of a tendency for network operators to have
focused efforts on maintaining infrastructure for censoring {\tt A} queries
sent to IPv4 resolvers, while paying less attention to the handling of {\tt
AAAA} queries and their IPv6 interfaces. It also presents an opportunity for
circumvention tool developers to exploit.
China presents the only exception with a preference for blocking
{\tt AAAA} queries on both IPv4 and IPv6 interfaces of resolvers with a 10\%
and 13\% higher {\tt AAAA} censorship rate, respectively. We investigate this
anomaly in \Cref{sec:cases}.

\begin{table}[t]
  \centering
  \small
  \scalebox{\tabularscale} {
  \begin{tabular}{lccc}%
    \toprule
    {\bf Country}&{\bf IPv4 resolvers}&{\bf IPv6 resolvers} & {\bf All resolvers}
    \\ \midrule
    Thailand (TH)      & -7.1 pp (-85.7\%) & $ns$               & -3.7 pp (-77.5\%) \\
    Bangladesh (BD)    & -5.1 pp (-80.0\%) & $ns$               & -2.6 pp (-71.3\%) \\
    Pakistan (PK)      & -2.1 pp (-73.6\%) & -2.8 pp (-59.8\%)  & -2.5 pp (-60.2\%) \\
    Chile (CL)         & -2.0 pp (-57.6\%) & $ns$               & -1.1 pp (-58.9\%) \\
    Vietnam (VN)       & $ns$              & -0.7 pp (-52.5\%)  & -0.7 pp (-47.1\%) \\
    Korea (KR)         & -1.3 pp (-54.6\%) & $ns$               & -0.6 pp (-36.2\%) \\
    China (CN)         &  3.1 pp (+10.4\%) &  3.7 pp (+12.9\%)  &  3.4 pp (+11.7\%) \\
    \midrule
    United States (US) & -0.5 pp (-33.2\%) &  0.4 pp (+72.3\%)  &  $ns$  \\
    Myanmar (MY)       & -2.9 pp (-58.9\%) & $ns$    &  $ns$  \\
    \bottomrule
  \end{tabular}
  }
  \caption{Differences in blocking rates of {\tt A} and {\tt AAAA} queries
  observed over IPv4, IPv6, and all resolvers in a country. `pp' denotes the
  change in terms of percentage points (computed as {\tt AAAA} blocking rate
  - {\tt A} blocking rate) and the \%age value denotes the percentage change in
  blocking rate (computed as
  $
  100\times\frac{\text{{\tt AAAA} blocking rate} - \text{{\tt A} blocking rate}}
  {\text{{\tt A} blocking rate}}
  $). 
  Only countries having a statistically significant difference are reported. A
  negative value indicates that {\tt A} queries observed higher blocking rates
  than {\tt AAAA} queries in a given country. $ns$ indicates the difference was
  not statistically significant and thus omitted. The United States and Myanmar
  are included separately as they show significance between IPv4 and IPv6
  censorship (\Cref{sec:infrastructure}).}
  \label{tab:resources:countries}
\end{table}

\subsection{{\tt A/AAAA}-inconsistent resolvers}
\label{sec:resources:resolvers}
Given our above results which suggest that there are a number of countries in
which {\tt A} and {\tt AAAA} queries are censored differently, we now seek to
understand the characteristics of the resolvers that cause these differences.
We first focus on identifying the individual resolvers in each country that
have statistically different behaviors for {\tt A} and {\tt AAAA} queries.
Then, we compare the AS distributions of these resolvers with the set of all
resolvers in a country. This comparison tells us if the {\tt A}/{\tt AAAA}
inconsistencies are specific to a subset of ISPs, or if the inconsistencies
exist across the whole country (suggesting centralized censorship).
Finally, we identify the
types of networks hosting inconsistent resolvers to get a measure of whether
users in residential networks may exploit these DNS inconsistencies for
circumvention.

\para{Identifying differences in individual resolvers.}
We begin our analysis by identifying the individual resolver pairs (\ie we
consider the IPv4- and IPv6-interfaces of a resolver as a unit), within each
of the seven countries listed above that have a statistically significant
difference in their censorship of {\tt A} and {\tt AAAA} queries.
To measure differences in DNS query handling of individual resolvers, we
compare the ratio of censored responses each resolver observes for {\tt A} and
{\tt AAAA} queries.

We use a two-proportion $z$-test to verify the statistical significance of any
observed difference in the ratios between the two groups for each resolver.
Similar to our within-country analysis, we apply a \Sidak correction to account
for our testing of multiple hypotheses and use $p \leq 1-{.05}^{1/n_{r_c}}$ to
classify a difference as significant, where $n_{r_c}$ is the total number of
resolvers in our dataset belonging to country $c$.
A summary of our results is provided in \Cref{tab:resources:resolvers}. 

\para{Which countries have the largest fractions of resolvers
exhibiting {\tt A} and {\tt AAAA} resolution inconsistencies?}
Immediately standing out from the other countries are Thailand, Bangladesh, and
Pakistan. These countries have {\tt A} and {\tt AAAA} inconsistencies in
62-82\% of their resolvers. In comparison, other countries with
statistically significant differences have inconsistencies arising from
anywhere between 3-30\% of their resolvers.

\para{How spread out are the {\tt A/AAAA}-inconsistent resolvers?} 
We calculate the entropy of the distribution of censorship by record query type
of all resolvers in the country ($S_{\text{query}}^{\text{all}}$) and compare it
with the entropy of the distribution of censorship by record query type of the
inconsistent resolvers in the country
($S_{\text{query}}^{\text{inconsistent}}$). This serves as a measure of the
diversity of ASes observed in both cases. 
In order to compare the two measures, we use the Kullback-Leibler divergence
($\nabla_{\text{query}}$) distance \cite{KLdivergence}. In simple terms, the
KL-divergence between two distributions ($X$, $Y$) measures the number
of additional bits required to encode $X$ given the optimal encoding for $Y$.
In other words, it is the relative entropy of one distribution given another. 
This computation is helpful for hypothesizing the censorship infrastructure
that causes the inconsistencies. Finding a small $\nabla_{\text{query}}$ value
in a country signifies that the inconsistent resolvers had a similar
distribution to all the resolvers in that country. This would suggest the
presence of a centralized mechanism that (roughly) equally impacts all ASes in
the country that is responsible for the inconsistencies.
Conversely, a higher $\nabla_{\text{query}}$ value indicates that there is
a strong change in the distribution of resolvers -- \ie a disproportionate
number of inconsistencies arise from a smaller set of ASes. This would be
indicative of local configuration inconsistency (at the network or resolver
level), rather than a centralized configuration inconsistency.
We note that this does not provide a measure of how centralized censorship is
within a country overall, but rather, it describes how uniform the {\tt A} vs.
{\tt AAAA} inconsistencies are.

Based on this analysis, we once again see that Thailand, Bangladesh, and
Pakistan stand out with small $\nabla_{\text{query}}$ values (0.14 - 0.48).
This suggests a country-wide censorship mechanism is responsible
for the inconsistent censorship of {\tt A} and {\tt AAAA} that impacts all
ASes nearly equally.
Korea, China, and the United States on the other hand demonstrate high
$\nabla_{\text{query}}$ scores suggesting the presence of network- or
resolver-level misconfigurations are responsible. This is confirmed by inspecting the ASes
hosting the resolvers with inconsistencies. For example, in the United States,
resolvers in just 5 ASes (of 249 ASes with resolvers) account for 56\% of all
{\tt A} and {\tt AAAA} inconsistencies.

\begin{table*}[t]
  \centering
  \small
  \scalebox{\tabularscale} {
    \begin{tabular}{lcclcccl}
    \toprule
      {\bf Country} & {\bf Total pairs} & {\bf Inconsistent pairs} & {\bf Most inconsistent AS} & \multicolumn{3}{c}{\bf AS diversity} & {\bf Most inconsistent type} \\ 
      & & {(\% of total pairs)}& {(\# inconsistent pairs)} & $S^{\text{all}}_{\text{query}}$ & $S^{\text{inconsistent}}_{\text{query}}$ & $\nabla_{\text{query}}$  & {(\# inconsistent pairs)} \\
      \midrule
      Thailand (TH)       & 186 & 152 (81.7\%)  &  AS9835 Government IT Services (40)  & 4.50 & 4.06 & 0.14 & Cable/DSL (110) \\
      Bangladesh (BD)     & 29  & 18 (62.1\%) & AS 9230 Bangladesh Online (4)          & 4.10 & 3.61 & 0.48 & Cable/DSL (18) \\    
      Pakistan (PK)       & 23  & 15 (65.2\%) & AS 17911 Brain Telecom (3)             & 3.43 & 3.06 & 0.25 & Cable/DSL (12) \\    
      Chile (CL)          & 65  & 20 (30.1\%) & AS 27651 Entel Chile (13)              & 3.08 & 1.14 & 1.18 & Corporate (13) \\    
      Vietnam (VN)        & 252 & 64 (25.4\%) & AS 131353 NhanHoa Software (37)        & 3.89 & 2.22 & 0.71 & Cable/DSL (59) \\    
      Korea (KR)          & 632 & 80 (12.7\%) & AS 9848 Sejong Telecom (13)            & 3.12 & 4.17 & 1.30 & Cable/DSL (58) \\    
      China (CN)          & 194 & 6 (3.1\%)   & AS 4538 China Education and Research Network (2) & 3.89 & 2.25 & 2.56 & Corporate (4)   \\    
    \midrule
      United States (US)  & 1,228 & 175 (14.3\%)  & AS 30475 WEHOSTWEBSITES (35) & 6.28 & 4.69 & 1.31 & Corporate (129) \\    
      Myanmar (MY)        & 50  & 30 (60\%)   & AS 136170 Exabytes Network (10)  & 3.31 & 2.42 & 0.48 & Corporate (26) \\    
    \bottomrule
  \end{tabular}
  }
  \caption{Characteristics of the resolvers which demonstrated a statistically
  significant difference in their handling of {\tt A} and {\tt AAAA} queries in
  each country. 
  `AS diversity' denotes the entropies of (all) resolver distribution
  ($S^{\text{all}}_{\text{query}}$) and {\tt A/AAAA}-inconsistent resolver
  distribution ($S^{\text{inconsistent}}_{\text{query}}$) across a country's
  ASes, and `$\nabla_{\text{query}}$' represents the Kullback-Leibler
  divergence of the distribution of inconsistent resolvers from the
  distribution of all resolvers in the country's ASes (\cf
  \Cref{sec:resources:resolvers}).
  `Most inconsistent type' denotes the connection type with the most number of
  {\tt A/AAAA}-inconsistent resolvers.The United States and Myanmar are included
  separately as they show significance between IPv4 and IPv6 censorship
  (\Cref{sec:infrastructure}).}
  \label{tab:resources:resolvers}
\end{table*}

\para{What types of networks exhibit the most {\tt A} and {\tt AAAA}
inconsistencies?}
We use the Maxmind GeoIP2 connection type database (retrieved in 01/2022
\cite{maxmind-connectiondb}) to identify the connection type of the resolvers
responsible for {\tt A} and {\tt AAAA} inconsistencies. 
We find that, in most countries, Cable/DSL network connections (typically
associated with residential networks) were most likely to host a resolver
exhibiting an inconsistency. 
Of the seven countries with statistically significant overall differences, only
Thailand and China were found to have a high ratio of {\tt A/AAAA} -inconsistent
resolvers in corporate networks. 
Combined with our previous results which suggest the presence of an
inconsistency in a centralized mechanism in Thailand, Bangladesh, and Pakistan,
these results show that these inconsistencies are likely extending to
residential networks --- a promising sign for the citizen users of
circumvention tools which exploit the {\tt A/AAAA} gap.

\subsection{Characteristics of anomalous domains} 
\label{sec:resources:domains}
We now identify the category of domains that appear to exist in the gap of {\tt
A} and {\tt AAAA} blocking. In addition to providing a general categorization
of these domains, we also analyze whether their category distributions vary
significantly from the category distribution of websites that received any
blocking. We do this in order to identify the specific policies or mechanisms
that differ between the censorship mechanisms for {\tt A} and {\tt AAAA}
queries.

\para{Identifying differences in domain behaviors within a country.} 
We continue using our statistical approach for identifying differences within
a country. We measure the ratio of blocking that occurs for a domain's {\tt A}
and {\tt AAAA} records within the country and then compare these using
a two-proportion $z$-test with a \Sidak corrected $p$-value of $1-.05^{dom}$
where $dom$ is the total number of tested domains (714). We only label the
behavior of a censor with regards to a domain as different over {\tt A} and
{\tt AAAA} queries if the $z$-test finds the difference to be statistically
significant. Once again, we do this to err on the side of caution in order to
minimize over-reporting and false-positives of censorship and, in this case,
its corresponding policy differences over {\tt A} and {\tt AAAA}.

\para{Do these {\tt A/AAAA}-inconsistent domains hint at policy gaps?}
For each country, we begin our analysis by deriving domain categories, using
the McAfee domain categorization service \cite{mcafee}, for domains in the
following two lists: (1) $\text{D}_{\text{any}}$ which contains
the domains which experienced any blocking events inside a country and (2)
$\text{D}_{\text{inconsistent}}$ which contains the {\tt
A/AAAA}-inconsistent domains identified by our $z$-test. Next, we compute the
KL-divergence between the category distributions of the two lists
(\ie $\nabla_{\text{query}}^{\text{domains}}
= \text{KLDivergence}(\text{D}_{\text{any}},
\text{D}_{\text{inconsistent}})$). 
A small $\nabla_{\text{query}}^{\text{domains}}$ would signify that the domains
that experience inconsistent treatment are not from a largely different
category distribution than the set of all domains that experience any type of
blocking. This would suggest that the inconsistencies do not arise from
a content-specific policy gap that exists in the censorship mechanism
implemented over {\tt A} and {\tt AAAA} queries. On the other hand, a large
difference would signify that the category distributions are very different and
that domains with specific types of content appear to have more
inconsistencies --- suggesting a content-based policy gap.

Based on this analysis, we find that the United States and Thailand have the
smallest $\nabla_{\text{query}}^{\text{domains}}$ scores (0.4-1.2). Conversely,
Pakistan and Myanmar have high $\nabla_{\text{query}}^{\text{domains}}$
scores (>2). 
After manual inspection of these results, we attribute the low divergence in
the United States to the fact that censorship observed in the US arise largely
from a range of corporate networks (\Cref{sec:prevalence}). These resolvers are fairly
consistent in their blocking of domains belonging to McAfee's `P2P/File Sharing' 
,`Malicious Sites' and `Technical/Business Forums' categories. Of the most blocked
individual domains, we see domains related to cryptocurrency (which were categorized
under `Technical/Business Forums'), `netflix.com' and `utorrent.com'. 
On the other hand, Thailand shows low divergence despite the large
number of residential networks. This suggests that there is no significant
policy gap that causes the {\tt A/AAAA}-inconsistencies. Rather, it suggests
an incomplete implementation of an existing mechanisms for {\tt A} query
censorship. 
On the other hand, in Pakistan and Myanmar where the divergence scores were
high, we found that the biggest contribution to the high divergence scores
arose from the `Pornography' and `Government/Military' domain categories,
respectively. This suggests the presence of a content-policy gap in the
implementation of {\tt A/AAAA} DNS censorship implementations that
disproportionately allows sites in these categories to evade censorship.

\section{Censorship of IPv4 and IPv6 Resolvers}
\label{sec:infrastructure}

\para{Overview.} In this section, we focus on the difference in censorship
for DNS queries sent to the \textbf{IPv4 and IPv6 interfaes of resolvers}.
Specifically, we answer the following questions:
(\Cref{sec:infrastructure:country}) In which countries are the DNS censorship
mechanisms for IPv4 and IPv6 traffic significantly different?,
(\Cref{sec:infrastructure:resolvers}) what are the characteristics of resolvers
that exhibit differences in the censorship of IPv4 and IPv6 traffic?, and
(\Cref{sec:infrastructure:domains}) what are the characteristics of the domains
in which such differences are frequently exhibited?

\subsection{IPv4 vs IPv6 infrastructure censorship} \label{sec:infrastructure:country}

To measure differences in censorship of DNS queries sent over IPv4 and IPv6 we
examine distributions similar to \Cref{sec:resources:country} but this time
investigate two distributions (corresponding to the IPv4 and IPv6 interfaces of
resolvers) of the fraction of censored domains from resolvers within the
corresponding country.
We again use a two-sample $t$-test to with \Sidak correction ($p \leq
1-{.05}^{1/n_{c}}$, where $n_c$ is
the total number of countries in our dataset (106)) to verify statistical
significance of any observed differences.
The presence of a statistically significant difference for a specific country
would imply that the country appears to have different censorship mechanisms
for IPv4 and IPv6 DNS traffic (if a centralized mechanism for censorship
exists) or that a significant number of resolvers within that country are not
consistent in their censorship of IPv6 and IPv4 traffic.
A summary of our results are presented in \Cref{tab:infrastructure:countries}.

\para{Which countries demonstrate large-scale inconsistencies in their
handling of IPv4 and IPv6 DNS traffic?}
In total, we find only five countries (Thailand, Iran, Bangladesh, Myanmar, and
the United States) with statistically significant differences in their handling
of DNS queries over IPv4 and IPv6 traffic --- suggesting the use of independent
censorship mechanisms for IPv4 and IPv6. 
Interestingly, all these countries appear to have gaps in their IPv6 censorship
apparatus --- \ie IPv4 rates of blocking are higher than IPv6 rates in all
countries with significant differences. These differences result in IPv6
queries experiencing between 12\% and 78\% less censorship than their IPv4
counterparts.
Once again, this suggests a tendency for network operators to more effectively
maintain IPv4 DNS censorship infrastructure than IPv6 infrastructure. These
gaps present opportunities for the success of circumvention tools with IPv6
capabilities.  
Further analysis shows that these differences primarily apply to {\tt A}
queries, suggesting that these countries fail to censor most {\tt A} queries
when they are sent over IPv6, but that they are able to censor
{\tt AAAA} records sent over IPv6 effectively.
These findings are particularly noteworthy for circumvention
efforts in Thailand, Myanmar, and Iran where IPv6 adoption rates are high
(between 15\% and 45\%) and dual-stack tools may be used for circumvention of
DNS censorship.

\begin{table}[t]
  \centering
  \small
  \scalebox{\tabularscale} {
  \begin{tabular}{lccc}%
    \toprule
    {\bf Country}&{\bf {\tt A} queries }&{\bf {\tt AAAA} queries} & {\bf All queries}
    \\ \midrule
    Thailand (TH)      & -7.1 pp (-86.3\%) & $ns$              & -3.7 pp (-78.1\%) \\
    Iran (IR)          & -3.2 pp (-12.6\%) & -3.0 pp (-12.5\%) & -3.1 pp (-12.5\%) \\ 
    Bangladesh (BD)    & -5.5 pp (-86.1\%) & $ns$              & -3.0 pp (-77.9\%) \\
    Myanmar (MY)       & -3.6 pp (-74.2\%) & $ns$              & -2.1 pp (-62.3\%) \\
    United States (US) & -0.9 pp (-64.8\%) & $ns$              & -0.5 pp (-42.6\%) \\
    \midrule
    Korea (KR)         & -1.1pp (-46.5\%) & $ns$    & $ns$ \\
    Chile (CL)         & -1.6pp (-58.7\%)  & $ns$    & $ns$ \\
    \bottomrule
  \end{tabular}
  }
  \caption{Differences in blocking rates of DNS queries sent to IPv4 and IPv6
  interfaces of each resolver in a country. `pp' denotes the change in
  terms of percentage points (computed as blocking rate of IPv6 - blocking
  rate of IPv4) and the \%age value denotes the percentage change in blocking rate
  (computed as 
  $
  100 \times \frac{\text{IPv6 blocking rate} - \text{IPv4 blocking rate}}
  {\text{IPv4 blocking rate}}
  $). 
  Only countries having a statistically significant difference are reported. A
  negative value indicates that queries sent over IPv4 observed higher blocking
  rates than those sent over IPv6 in a given country. $ns$ indicates the
  difference was not statistically significant and thus omitted. Korea and Chile
  are included separately as they demonstrated significant difference in
  \texttt{A}/\texttt{AAAA} censorship (\Cref{sec:resources:country}).}
  \label{tab:infrastructure:countries}
\end{table}

\subsection{IPv4/IPv6-inconsistent resolvers}
\label{sec:infrastructure:resolvers}

Next, we look at the characteristics of resolver pairs that see inconsistent
censorship on their IPv4 and IPv6 interfaces. 

\para{Identifying IPv4/IPv6-inconsistent resolvers.} Our approach is similar to
the methods used to identify {\tt A/AAAA}-inconsistent resolvers
(\cf \Cref{sec:resources:resolvers}). 
We compare the ratios of censored responses received from a single resolver
pair's IPv4 and IPv6 interfaces. We test whether these ratios are statistically
different using a $z$-test with a \Sidak corrected $p \leq 1-.05^{1/n_{r_c}}$
being required for a statistically significant difference. 
A summary of the characteristics of the inconsistent resolvers identified in
each country is illustrated in \Cref{tab:infrastructure:resolvers}.

\para{Which countries have the largest fraction of IPv4/IPv6-inconsistent
resolvers?}
Two countries from our previous analysis on {\tt A/AAAA}-inconsistencies once
again appear with a large fraction of IPv4/IPv6-inconsistent resolvers ---
Thailand (81\%) and Bangladesh (65\%). Myanmar presents a new addition with
60\% of its resolvers demonstrating IPv4/IPv6-inconsistencies. Other countries
were found to have smaller fractions ranging from 12-26\%.

\para{How spread out are the IPv4/IPv6-inconsistent resolvers?}
In order to characterize the spread of IPv4/IPv6-inconsistent resolvers within
a country, we compute the entropy of the AS distribution of all resolvers and
IPv4/IPv6-inconsistent resolvers within a country
($S^{\text{all}}_{\text{net}}$ and $S^{\text{inconsistent}}_{\text{net}}$) and
then compute the KL-divergence of the distribution of inconsistent resolvers
from the distribution of all resolvers in that country ($\nabla_{net}$).
Similar to before, a large change in $\nabla_{net}$ means that the
IPv4/IPv6-inconsistencies arise from a small fraction of ASes and would suggest
that the gaps exist due to local network/resolver configurations --- as
would be the case if regional operators implement their own DNS censorship
mechanisms. Conversely, a small change means that the gaps that exist roughly
equally impact all the ASes having resolvers and would suggest that the gaps
exist due to misconfigurations in a centralized DNS censorship mechanism.
Our results again suggest the presence of a centralized blocking mechanism
in Thailand, Bangladesh, and Myanmar ($\nabla_{net} \in [0.13, 0.48]$) which
causes the IPv4/IPv6-inconsistencies. The United States has the highest
$\nabla_{net}$ observed which indicates that regional policies are responsible
for the IPv4/IPv6-inconsistencies. 

\para{What types of networks exhibit the most IPv4 and IPv6 inconsistencies?}
An overwhelming majority of the inconsistent resolvers in Thailand, Iran, and
Bangladesh (77\%-100\%) are found to be present in networks with (Maxmind
categorized) Cable/DSL connection-types that are typically associated with
residential networks. 
Put in the context of our previous result which suggests the presence of
a centralized DNS censorship mechanism in Thailand and Bangladesh, this
suggests that the IPv4/IPv6 gaps that exist in this mechanism also extend to
residential networks in the country.
Myanmar and the United States experience such inconsistencies primarily due to
their corporate networks which contain between 67-87\% of their inconsistent
resolvers. 

\begin{table*}[t]
  \centering
  \small
  \scalebox{\tabularscale} {
    \begin{tabular}{lcclcccl}
    \toprule
      {\bf Country} & {\bf Total pairs} & {\bf Inconsistent pairs} & {\bf Most inconsistent AS} & \multicolumn{3}{c}{\bf AS diversity} & {\bf Most inconsistent type} \\
      & & {(\% of total pairs)}& {(\# inconsistent pairs)}
      & $S^{\text{all}}_{\text{net}}$ & $S^{\text{inconsistent}}_{\text{net}}$
      & $\nabla_{\text{net}}$  & {(\# inconsistent pairs)} \\
      \midrule
      Thailand (TH)            & 186    & 151 (81.2\%)  & AS 9835 Government IT Services (39)  & 4.50 & 4.10 & 0.13 & Cable/DSL (108) \\
      Iran (IR)                & 277    & 74 (26.7\%)   & AS 208161 PARSVDS (11)               & 5.03 & 3.81 & 0.87 & Cable/DSL (57) \\
      Bangladesh (BD)          & 29     & 19 (65.2\%)   & AS 9230 Bangladesh Online (4)        & 4.10 & 3.72 & 0.39 & Cable/DSL (19) \\
      Myanmar (MY)             & 50     & 30 (60.0\%)   & AS 136170 Exabytes Network (10)      & 3.31 & 2.42 & 0.48 & Corporate (26) \\
      United States (US)       & 1,228  & 151 (12.3\%)  & AS 30457 WEHOSTWEBSITES (36)         & 6.28 & 5.22 & 1.44 & Corporate (102) \\
      \midrule                 
      Korea (KR)               & 632    & 101 (16.0\%)  & AS 9848 Sejong Telecom (13)          & 3.12 & 4.14 & 0.95 & Cable/DSL (73)   \\
      Chile (CL)               & 65     & 16 (24.6\%)   & AS 27651 Entel Chile (12)            & 3.08 & 1.19 & 1.04 & Corporate (11) \\
    \bottomrule
  \end{tabular}
  }
  \caption{Characteristics of the resolvers which demonstrated a statistically
  significant difference in their handling of DNS queries over IPv4 and IPv6
  each country.
  `AS diversity' denotes the entropies of (all) resolver distribution
  ($S^{\text{all}}_{\text{net}}$) and {IPv4/IPv6}-inconsistent resolver
  distribution ($S^{\text{inconsistent}}_{\text{net}}$) across a country's
  ASes, and `$\nabla_{\text{net}}$' represents the Kullback-Leibler divergence
  of the distribution of inconsistent resolvers from the distribution of all
  resolvers in the country's ASes (\cf \Cref{sec:infrastructure:resolvers}).
  `Most inconsistent type' denotes the connection type with the most number of
  {IPv4/IPv6}-inconsistent resolvers.Korea and Chile are included separately as
  they demonstrated significant difference in \texttt{A}/\texttt{AAAA}
  censorship (\Cref{sec:resources:country}).}
  \label{tab:infrastructure:resolvers}
\end{table*}

\subsection{Characterization of anomalous domains}
\label{sec:infrastructure:domains}
We now seek to understand the category of domains that get through the
infrastructural gap between DNS queries over IPv4 and IPv6. In other words,
does the category of domain influence whether or not there is a difference in
censorship between IPv4 and IPv6 interfaces?

\para{Identifying differences in domain behaviors within a country}
We use an approach similar to the one defined in \Cref{sec:resources:domains}
for identifying domains that get through the gap in between IPv4 and IPv6. For
each domain, we measure the ratio of blocking that occurs over IPv4 and IPv6. We
then apply the $z$-test with a corrected $p$-value (as described in
\Cref{sec:resources:domains}) to these ratios. This gives us all domains which
had significant differences in censorship over IPv4 and IPv6.

\para{Do these IPv4/IPv6-inconsistent domains hint at policy gaps?}
For each country, we derive domain categories for the following two lists; (1)
$\text{D}_{\text{any}}$ which contains the domains which experienced any
blocking events inside a country and (2) $\text{D}_{\text{inconsistent}}$ which
contains the IPv4- and IPv6-inconsistent domains identified by our $z$-test. We
again used KL-divergence between to compare these two category distributions. A
small $\nabla_{\text{net}}^{\text{domains}}$ would signify that the two
category distributions are not largely different suggesting the policy gap over
IPv4/IPv6 is not content specific; A large
$\nabla_{\text{net}}^{\text{domains}}$ would signify that the category
distributions are indeed very different suggesting a content-based policy gap.
\\
Based on this analysis, we see that Bangladesh, US and Iran have the smallest
$\nabla_{\text{net}}^{\text{domains}}$ scores (0-1.3) suggesting that there
was little to no difference in the distribution of the two categories. This
suggests that for these countries, there is no content-based policy gaps. On the
other hand, Pakistan, China and Myanmar had the highest
$\nabla_{\text{net}}^{\text{domains}}$ scores (2.7-4.3). These scores suggest
that some categories of domains get through the IPv4/IPv6 censorship gap much
more than others. "Gambling" is the category most likely to get through this gap
for China and Pakistan and "Government/Military" is the most likely to get
through this gap in Iran. 

\section{DNS Censorship Case studies}
\label{sec:cases}
In this section, we provide interesting findings about the censorship
infrastructure in three heavily censored regions which have high IPv6 adoption
and have not been discussed in detail in prior sections: Iran, Thailand, and
China.

\IranRecursionTable

\para{Iran.}
While Iran appears to support censorship of DNS queries in both IPv4 and IPv6,
it is more effective at blocking queries from IPv4 resolvers (in both \texttt{A}
and \texttt{AAAA} records). Notably, 272 of the 277 IPv6 resolvers in Iran rely
on 6to4 bridges, all of which block queries over IPv6 at statistically
significantly lower rates (-3.2\%) than queries over IPv4.

Given the prevalence of 6to4 resolvers in Iran we performed a manual follow up
investigation to better understand their censorship dynamics. We sent test
queries from a native IPv6 host outside of the country to resolvers within the
country and found that injected responses have discernible characteristics. The
injected response always contains a single answer record with the same address
independent of the censored domain or the requested record type in the
triggering query. Requests sent over IPv4 receive an \texttt{A} record with the
address \texttt{10.10.34.35} while requests sent to the (very few) native IPv6
resolvers receive an \texttt{AAAA} record with \texttt{d0::11}. However, as
shown in Table~\ref{tab:iran-recursion-table}, this rule does not follow for
queries to 6to4 resolvers which receive the v4 injection normally and no
injection when the recursion desired flag is unset. This indicates that DNS
packets encapsulated in 6to4 are not parsed by the censor, but recursive lookups
done by the resolvers in the country do receive injected responses. This
explains the significant similarity of 6to4 and IPv4 censorship rates in the
country as forwarded or recursive lookups for nameservers associated with
censored domains are themselves triggering censorship responses in IPv4 that are
then passed on to the querying client.

\para{Thailand.}
In Thailand, we see significant differences in the handling of {\tt A} queries
sent over IPv4 and all other query-network combinations --- 8.2\% of \texttt{A}
queries sent to IPv4 resolvers were censored, compared to $\approx$ 1\% of all
other query and network combinations.
For instance, examining the IPv4 interface of one of the resolvers with the
largest blocked domains list: it blocked 240 \texttt{A} records, but only 15
\texttt{AAAA} records from our domain list. Meanwhile, its IPv6 counterpart did
not block any domains (either \texttt{A} or \texttt{AAAA} records). These
findings suggest an incomplete DNS censorship infrastructure that is yet to
effectively handle IPv6 traffic.
Furthermore, the number of requests censored broken down by connection type
closely matches the proportion of the resolvers of that connection type. For
example, ~74\% of all resolvers in Thailand are of type Cable/DSL and they are
responsible for ~72\% of all blocked queries. This, coupled with findings in
\Cref{sec:infrastructure:resolvers} (Thailand having a low $\nabla_{net}$)
present a strong case for a centralized mechanism for censorship.
Despite the already low rate of native IPv6 censorship (1.3\%), we find that
the 112 6to4-bridged resolvers had an even lower rate of censorship (0.6\%)
which suggests, once again, an inadequately equipped IPv6 censorship mechanism.

\para{China.}
China is unique in that it shows an unusual preference \textbf{toward} blocking
\texttt{AAAA} records over \texttt{A} records. While it is known that the Great
Firewall can block \texttt{AAAA} records and injects IPv6 traffic, it is not
clear why it would block those records more than \texttt{A} records. Manual
investigation reveals 21 domains that almost exclusively have their
\texttt{AAAA} record blocked, but not their \texttt{A} record. For example,
\texttt{gmail.com}'s \texttt{AAAA} record is blocked by over 95\% of resolvers
in China, but the corresponding \texttt{A} record for \texttt{gmail.com} is
only blocked by 1\% of resolvers. The other 20 domains have similar patterns,
leading to China's slight preference in blocking \texttt{AAAA} records over
\texttt{A} records. We do not find any instances of domains in China that are
similarly exclusively blocked by \texttt{A} record but not \texttt{AAAA}.
As discussed in \Cref{sec:infrastructure:domains}, China has a high
$\nabla_{net}$ which suggests that there are individual resolver-level
inconsistencies. 90\% of all resolvers in China fall within one standard
deviation of the mean number of censored queries by all resolvers in China.
Those that exceed these bounds are overwhelmingly classified as `Cellular' by
Maxmind's connection type database. Although they make up just 17\% of all our
resolvers in China, they present 50\% of the resolvers with censorship rates
higher than mean+std.
China has a much larger percentage of native IPv6 resolvers compared to our
other case studies and it is still the most consistent in its censorship rate
for \texttt{AAAA} queries over both 6to4 and native IPv6 resolvers (~30\%) when
compared with their IPv4 counterparts. However, the censorship rate for 6to4
when compared with native IPv6 resolvers was still slightly lower (28\% vs
32\%).
Taken all together, China appears to be the best equipped IPv6 censor with gaps
only arising in the IPv4 censorship apparatus.

\section{Related Work}\label{sec:related}
There is a significant body of previous work investigating DNS based censorship
strategies which can be generally broken down along several dimensions: by their
duration, scope, and vantage points.

\para{Longitudinal or snapshot measurements of DNS censorship.}
Many initial investigations into censorship techniques and
proposals for measurement methodology provide an evaluation of DNS censorship
during a single snapshot in time \cite{Anonymous2020:TripletCensors,
global2002great, vandersloot2018quack, scott2016satellite, pearce2017global}. 
Similarly, a snapshot can capture DNS censorship centering around a specific
event like an election or social uprising \cite{aryan2013internet}.  Established
methodology can then be used to gain perspectives on the way that censorship
evolves over time as part of  a longitudinal measurement
\cite{USESEC21:GFWatch, filasto2012ooni, sundara2020censored, niaki2020iclab,
razaghpanah2016exploring}. Our work provides a snapshot view of DNS censorship
during the period of transition from IPv4 to IPv6.

\para{Regional or global measurements of DNS censorship.}
Censorship strategies are not universal and each censor is unique to some
degree. Targeted measurement studies contribute to a better understanding of
block-list infrastructure ~\cite{ramesh2020decentralized, USESEC21:GFWatch} and
explain blocking phenomena~\cite{global2002great,
Anonymous2020:TripletCensors}. Global studies provide high level view on the
use of DNS censorship internationally providing context and and understanding
of prevalence to specific censorship techniques~\cite{vandersloot2018quack,
scott2016satellite, pearce2017global, sundara2020censored, niaki2020iclab}.
Our work performs a global measurement of DNS censorship, however, through the
lens of the differences in IPv4 and IPv6 censorship deployment, we are able to
make data-driven hypotheses about the censorship infrastructure deployed in
several countries.

\para{Using open resolvers as measurement vantage points.}
Open DNS resolvers provide a unique vantage point from which to study the
Internet and a significant number of previous efforts have relied on them to
gain insights into the censorship mechanisms in different regions.

In 2007, Lowe \etal queried open resolvers hosted within China eliciting
injected responses. This was the first use of open resolvers to characterize
Chinese censorship infrastructure and strategy~\cite{lowe2007great}. In 2008
In 2012, anonymous authors focused on the Chinese Great Firewall's DNS
injection and the collateral poisoning effect that it had on open resolvers
around the world~\cite{nebuchadnezzar2012collateral}. In 2014, Wander \etal
used open resolvers to look more broadly for global poisoning of DNS resolution
by any censoring country finding that spoofed addresses were leaking primarily
from China and Iran ~\cite{wander2014measurement}. Similar to Dagon \etal, in
2015 K{\"u}hrer \etal performed a global study of the reliability of open DNS
resolution finding evidence of censorship, injected advertisements, and other
suspicious or malicious behavior by returned addresses~\cite{kuhrer2015going}.

More recently, Satellite (2016) outlined a methodology for regularly
discovering the set of available open resolvers and querying hosts in order to
detect paths that return incorrect or inconsistent resource
records~\cite{scott2016satellite}. Iris (2017) relied on a similar scanning
methodology but developed a set of metrics using follow-up scans and requests
that allow the differentiation between inconsistency, misconfiguration, and
manipulation~\cite{pearce2017global}. These metrics are based on several
factors including: address consistency, TLS certificate validation, HTTP
content hash, geolocation, DNS PTR lookup, and AS information. These
supplemental elements allow Iris to handle cases like CDNs, virtual hosting,
distributed / forwarded resolution requests and more. The 2020 Censored Planet
project incorporated and extended the methods of the Satellite and Iris as part
of a comprehensive and longitudinal global censorship measurement
study~\cite{sundara2020censored}. 
Although several of these measurement platforms perform filtering to identify
reliable resolvers, they are all limited to using open resolvers in the IPv4
space. Consequently, they are unable to provide insights into the state of IPv6
censorship.

\para{IPv6 censorship measurement.} 
Previous censorship studies primarily focus on measurements in the context of
IPv4. However, there have been several efforts to explicitly incorporate IPv6.
In March 2020 Hoang \etal began collection of DNS records injected by the Great
Firewall in order to classify the addresses provided, block-pages injected, and
the set of hostnames that receive injections~\cite{USESEC21:GFWatch}. Their
analysis investigates the commonality of addresses injected by the GFW, finding
that all injected \texttt{AAAA} responses are drawn from the reserved teredo
subnet \texttt{2001::/32}. However, because this study does not directly
compare the injection rates of A vs AAAA or differences in injection to DNS
queries sent over IPv4 versus IPv6, our efforts complement their findings and
provide a more detailed understanding of IPv6 censorship in China.
Although not focused on DNS censorship, a 2021 investigation of HTTP keyword
block-lists associated with the Great Firewall found that results are largely
the same between IPv4 and IPv6~\cite{weinberg2021chinese}. However, the authors
note that over IPv6 connections, the the Firewall failed to apply it's
signature temporary 90 second ``penalty box'' blocking subsequent connections
between the two hosts described by numerous previous
studies~\cite{xu2011internet,clayton2006ignoring}. This supports our finding
that for now the GFW infrastructure supporting IPv4 and IPv6 are implemented
and/or deployed independently.

\section{Discussion and Conclusions} \label{sec:discussion}

In this section, we detail limitations
(\Cref{sec:discussion:limitations}), directions for future research
(\Cref{sec:discussion:future}), and the takeaways of our work.
(\Cref{sec:discussion:conclusions}).

\subsection{Limitations}
\label{sec:discussion:limitations}
Measuring censorship in order to gain an understanding of the underlying
infrastructure and identify weaknesses for circumvention is a challenging task
due to the absence of ground truth for validation and the often probabilistic
nature of censorship and networking failures which are easily confused.
Although we take care to always err on the side of caution and consider many
confounding factors including end-point type and AS diversity, our work is
fundamentally a best-effort attempt at trying to identify the gaps that have
emerged in DNS censorship deployments because of the increased adoption of
IPv6. The limitations of our study arise from three sources.

\para{External sources of data.}
Our study relies on multiple data sources including Satellite and the Citizen
Lab for our domain lists, McAfee's domain labeling services for categorizing
our data, and Maxmind's datasets for geolocating and classifying connection
types of our resolvers. Although each of these datasets has been validated in
the past and are commonly used in research, our results and their corresponding
analyses are limited by their reliability.

\para{Statistical limitations.}
Throughout our study, we aimed to err on the side of caution in order to avoid
presenting false-positives in our results. Therefore, we relied on rigorous
statistical approaches in order to identify differences in censorship between
{\tt A/AAAA} query and {IPv4/IPv6} network types. This included grouping
related hypotheses together and performing Sidak corrections to ensure that the
confidence level achieved across the entire group (rather than for each
individual hypothesis) was 95\%. Although this provides our results with
credibility, our strict methods almost certainly assure false-negatives. In
fact, this becomes apparent in our identification of {\tt A/AAAA}-resolvers in
China. Our statistical methods identified only six resolvers with significant
differences (\Cref{sec:resources}), however, the manual analysis performed in
the case study (\Cref{sec:cases}) shows that a much larger number of resolvers
performed censorship over a small set of just 21 domains. This is a fundamental
limitation of any test of two proportions such as the $z$-test used in this
work.

\para{Resolver selection.}
Our work required us to identify resolvers that were IPv4 and IPv6 capable. For
this, we only partially followed the steps outlined by Hendricks \etal
(\cf \Cref{sec:methodology:resolvers}). We decided not to take explicit steps
to guarantee that each resolver pair was in fact a single dual stack resolver.
Doing so would have resulted in the complete removal of 6to4 bridges and
infrastructure resolvers. Our reasoning was that such resolvers do play an
important role in the censorship apparatus since they are the ones most
commonly used by citizens in a censored region. Therefore, we decided that
their removal would have harmed (1) the representativeness of our measurements
and (2) the effectiveness of any circumvention approaches that might seek to
exploit the censorship gaps identified in this work. On the other hand, this
choice means that it is possible that our dataset may have multiple pairs of
resolvers in a country that belong to the same resolving infrastructure. 

\parait{The presence of many 6to4 bridges.}
One specific category of open resolver that we find to be relatively common in
our collected data are those that rely on a {6to4 Bridge}. We find many
countries where IPv6 resolver availability is supplemented with 6to4 and note
that the filtering stage of our resolver methodology ensures that resolver IP
pairs geolocate within the same country. This means that any IPv6 queries sent
to 6to4 resolvers do traverse national networks as UDP over IPv6 for at least
some potion of their path to the final resolver. Further, if our queries do not
get served by the resolver's cache, it is expected that the recursive queries
will still pass through the censor. Therefore, still providing standard
censorship infrastructure some opportunities to censor.
Two specific examples of a substantial 6to4 supplement are Brazil in which 94
of the 160 resolvers identified used 6to4 and Thailand with 112 out of 186
resolvers. This understanding of the protocols that are traversing the censored
links provide a limitation on our measurement in the sense that we do not get
a perfect comparison of IPv4 with IPv6 that, however this trades-off with our
ability to gain an interesting interesting perspective on the ability of
censors to handle protocols specifically deployed to assist in bridging the
IPv4 - IPv6 gap. For example in both Brazil and Thailand the resolvers relying
on 6to4 censor requests at lower rates of injection for requests sent over IPv6
on 6to4 as compared with Native IPv6. This suggests an opportunity of
circumvention tool developers targeting citizens in these regions.

\subsection{Circumvention Opportunities}
\label{sec:discussion:future}
Our work identifies many IPv6-related DNS censorship inconsistencies around the
world. Each of these provides interesting directions for future work and
opportunities for circumvention tool developers.
One particularly interesting direction concerns the discrepancy in the
effectiveness of native IPv6 and 6to4-bridged IPv6 censorship. We find that
there are several countries where 6to4 bridges experience significantly lower
amounts of censorship --- including China and India. This presents the
possibility of simply using 6to4 tunnels to circumvent DNS censorship. 
Further, 6to4 is not the only protocol that has seen deployment as part of an
effort to bridge the gap between IPv4 and IPv6 deployment. The \texttt{6rd}
protocol extends the subnet range of 6to4. The Teredo protocol uses UDP to
encapsulate IP traffic. Finally, the ISATAP protocol provides an extended 6to4
adding configuration using multicast and DNS. This strategy is not only
applicable to DNS censorship circumvention as tunneling protocols can be used to
encapsulate any next layer.
Our work also highlights the need for integration of IPv6-related measurements
by longitudinal censorship measurement platforms such as Satellite, OONI, and
ICLab. 

\subsection{Conclusions} \label{sec:discussion:conclusions}

The current state of global IPv6 DNS censorship does not divide into a binary
bifurcation: block or allow, the landscape is nuanced and developing. However, we
identify clear trends in IPv6 censorship that can help motivate future research
and circumvention tool development.
In this work we present the first large scale measurement of DNS injection based
censorship in IPv6 and provide a reproducible methodology for incorporating this
measurement into longitudinal measurements going forward. We find significant
evidence of DNS injection in IPv6 in many countries around the world. Despite
this, we find that censoring countries still favor native network
infrastructure, censoring \texttt{A} queries over IPv4 at the highest rates. At
the country level
we observe discrepencies that inform if censorship is centrally coordinated,
or if there is looser coordination between
entities enacting the DNS censorship regimes.
The IP layer and the gap that exists between full IPv6 deployment and the
current state of the Internet provide a telling lens for evaluating censorship
and developing censorship circumvention technologies. IPv6 plays a growing role
in global connectivity and, wherever possible, censorship measurements should
endeavor to include IPv6 analysis to provide the most actionable understanding
of global censorship infrastructure.
 
\newpage
\bibliographystyle{plain}
\bibliography{censorship}

\newpage

\appendix

\section{Full list of countries / resolvers}

{\footnotesize

\bottomcaption{ Full list of measured rates of DNS censorship across both query
types and network interfaces. Rates are expressed as the percentage of censored
DNS queries over total number of DNS queries sent.
  Darker shaded cells indicate a higher rate of DNS censorship (compared to the
  country's average) and lighter shaded cells indicate a lower rate of DNS
  censorship.
  The average rate of censorship for a country is computed across all four
  IP/query combinations.
  The global row contains the mean of each column. These means weigh the
  contribution of each country equally, rather than weighted by the number of
  resolvers used in tests. \label{tab:appendix:rates}} 
\tablefirsthead{
  \toprule {\bf Country} &  {\bf Resolver} & {\bf IPv4} & {\bf IPv4} & {\bf IPv6} & 
  {\bf IPv6} & {\bf Avg.} \\ 
  {}  & {\bf pairs} & {\tt A} & {\tt AAAA}  & {\tt A} & {\tt AAAA} & {} \\ 
  \midrule } 
\begin{supertabular}{lcccccc} 
United States (US) & \color{black} 1228 & {\cellcolor[HTML]{6CAED6}} \color[HTML]{F1F1F1} \color{black} 1.40 & {\cellcolor[HTML]{C7DCEF}} \color[HTML]{000000} \color{black} 0.94 & {\cellcolor[HTML]{F7FBFF}} \color[HTML]{000000} \color{black} 0.49 & {\cellcolor[HTML]{D0E2F2}} \color[HTML]{000000} \color{black} 0.85 & \color{black} 0.92 \\
Germany (DE) & \color{black} 753 & {\cellcolor[HTML]{6CAED6}} \color[HTML]{F1F1F1} \color{black} 1.00 & {\cellcolor[HTML]{ADD0E6}} \color[HTML]{000000} \color{black} 0.92 & {\cellcolor[HTML]{F7FBFF}} \color[HTML]{000000} \color{black} 0.75 & {\cellcolor[HTML]{E0ECF8}} \color[HTML]{000000} \color{black} 0.81 & \color{black} 0.87 \\
Korea (KR) & \color{black} 632 & {\cellcolor[HTML]{6AAED6}} \color[HTML]{F1F1F1} \color{black} 2.33 & {\cellcolor[HTML]{F7FBFF}} \color[HTML]{000000} \color{black} 1.06 & {\cellcolor[HTML]{E8F1FA}} \color[HTML]{000000} \color{black} 1.24 & {\cellcolor[HTML]{EAF3FB}} \color[HTML]{000000} \color{black} 1.22 & \color{black} 1.46 \\
France (FR) & \color{black} 560 & {\cellcolor[HTML]{6CAED6}} \color[HTML]{F1F1F1} \color{black} 0.56 & {\cellcolor[HTML]{CCDFF1}} \color[HTML]{000000} \color{black} 0.47 & {\cellcolor[HTML]{F7FBFF}} \color[HTML]{000000} \color{black} 0.39 & {\cellcolor[HTML]{89BEDC}} \color[HTML]{000000} \color{black} 0.54 & \color{black} 0.49 \\
Russia (RU) & \color{black} 312 & {\cellcolor[HTML]{6AAED6}} \color[HTML]{F1F1F1} \color{black} 5.51 & {\cellcolor[HTML]{DCE9F6}} \color[HTML]{000000} \color{black} 4.78 & {\cellcolor[HTML]{F7FBFF}} \color[HTML]{000000} \color{black} 4.49 & {\cellcolor[HTML]{F7FBFF}} \color[HTML]{000000} \color{black} 4.50 & \color{black} 4.82 \\
\bottomrule
  \end{supertabular}
}
 \pagebreak
\section{Censorship rates by network end-point types}

{\footnotesize

\bottomcaption{Measured rates of censorship for countries with more than 25
resolver-pairs with corporate, non-corporate, native IPv6, and 6to4-bridged
IPv6 end-points. Rates are expressed as the percentage of censored DNS queries.
  Darker shaded cells indicate a higher rate of DNS censorship (compared to the
  country's average computed over all network and query types).
\label{tab:prevalence:rates:broken}}
\tablefirsthead{
  \toprule {\bf Country} &  {\bf Resolver} & {\bf IPv4} & {\bf IPv4} & {\bf IPv6} &
  {\bf IPv6} & {\bf Avg.} \\
  {}  & {\bf pairs} & {\tt A} & {\tt AAAA}  & {\tt A} & {\tt AAAA} & {} \\
  \midrule }
\begin{supertabular}{lcccccc}
   \midrule
    \multicolumn{7}{c}{Corporate censorship (>25 resolvers)}\\
    \midrule
Iran (IR) & \color{black} 48 & {\cellcolor[HTML]{6AAED6}} \color[HTML]{F1F1F1} \color{black} 25.50 & {\cellcolor[HTML]{99C7E0}} \color[HTML]{000000} \color{black} 24.66 & {\cellcolor[HTML]{E6F0F9}} \color[HTML]{000000} \color{black} 22.34 & {\cellcolor[HTML]{F7FBFF}} \color[HTML]{000000} \color{black} 21.69 & \color{black} 23.55 \\
Hong Kong (HK) & \color{black} 37 & {\cellcolor[HTML]{6AAED6}} \color[HTML]{F1F1F1} \color{black} 6.20 & {\cellcolor[HTML]{BFD8ED}} \color[HTML]{000000} \color{black} 5.44 & {\cellcolor[HTML]{F7FBFF}} \color[HTML]{000000} \color{black} 4.55 & {\cellcolor[HTML]{C6DBEF}} \color[HTML]{000000} \color{black} 5.38 & \color{black} 5.39 \\
Russian Federation (RU) & \color{black} 117 & {\cellcolor[HTML]{6AAED6}} \color[HTML]{F1F1F1} \color{black} 4.67 & {\cellcolor[HTML]{C1D9ED}} \color[HTML]{000000} \color{black} 4.36 & {\cellcolor[HTML]{F6FAFF}} \color[HTML]{000000} \color{black} 4.01 & {\cellcolor[HTML]{F7FBFF}} \color[HTML]{000000} \color{black} 4.00 & \color{black} 4.26 \\
Thailand (TH) & \color{black} 43 & {\cellcolor[HTML]{6AAED6}} \color[HTML]{F1F1F1} \color{black} 9.28 & {\cellcolor[HTML]{F5F9FE}} \color[HTML]{000000} \color{black} 1.18 & {\cellcolor[HTML]{F7FBFF}} \color[HTML]{000000} \color{black} 0.99 & {\cellcolor[HTML]{F7FBFF}} \color[HTML]{000000} \color{black} 0.95 & \color{black} 3.10 \\
Malaysia (MY) & \color{black} 30 & {\cellcolor[HTML]{6AAED6}} \color[HTML]{F1F1F1} \color{black} 6.07 & {\cellcolor[HTML]{DEEBF7}} \color[HTML]{000000} \color{black} 2.25 & {\cellcolor[HTML]{F7FBFF}} \color[HTML]{000000} \color{black} 0.97 & {\cellcolor[HTML]{F5F9FE}} \color[HTML]{000000} \color{black} 1.10 & \color{black} 2.60 \\
\bottomrule
  \end{supertabular}
  }

\end{document}